\documentclass[article]{aa}  
\usepackage{txfonts}
\usepackage{natbib}
\usepackage{graphicx}
\usepackage{color}
\usepackage{longtable,lscape}
\usepackage{float}
\usepackage{subfigure}
\newcommand{\kmpers}{$\mathrm{km \, s^{-1}}$} 

\newcommand{\cmtwo}{cm$^{-2}$} 
 
\newcommand{\cmthree}{cm$^{-3}$}

\newcommand{\kms}{km\,s$^{-1}$}               
\newcommand{\ecs}{erg cm$^{-2}$ s$^{-1}$}

\newcommand{\um}{$\mu$m}                      
\newcommand{\vlsr}{$\upsilon_{\rm LSR}$}              

\newcommand{\dv}{$d \upsilon$}  
\newcommand{\deltav}{$\Delta \upsilon$}
\newcommand{\vs}{$\upsilon_{\rm s}$}

\newcommand{\tmb}{$T_{\rm mb}$}

\newcommand{\tkin}{$T_{\rm kin}$}

\newcommand{\gapprox}{$\stackrel{>}{_{\sim}}$}     
\newcommand{\lapprox}{$\stackrel{<}{_{\sim}}$}

\newcommand{\about}{$\sim$}                       
\newcommand{\powten}[1]{10$^{#1}$}

\newcommand{\expo}[1]{$10^{#1}$}
\newcommand{\texpo}[1]{$\,\times\,10^{#1}$}

\newcommand{\av}{$A_{\rm V}$}                     

\newcommand{\ebv}{$E_{\rm B-V}$}

\newcommand{\Nhtva}{$N$($\mathrm{H_2}$)}           

\newcommand{\nhtva}{$n$($\mathrm{H_2}$)}           

\newcommand{\xhtvao}{$X$($\mathrm{H_2O}$)}            

\newcommand{\xohtvao}{$X(o$-$\mathrm{H_2O})$}
\newcommand{\xco}{$X$(CO)}

\newcommand{\ggr}{$\times$}
\newcommand{\orthowater}{\textit{ortho}-water}     
\newcommand{\molh}{H$_{2}$}  
\newcommand{\ohtvao}{\textit{o}-H$_2$O}

\newcommand{\phtva}{\textit{p}-H$_2$} 
\newcommand{\htvao}{H$_2$O}  
\newcommand{\htva}{H$_2$}

\newcommand{\halpha}{H$\alpha$}                   

\newcommand{\oishort}{[O\,{\sc i}]\,63\,$\mu$m}

\newcommand{\htvaoett}{$\rm{H_2O\,(1_{10}-1_{01})}$}
\newcommand{\htvaotva}{$\rm{H_2O\,(2_{12}-1_{01})}$}

\newcommand{\htvaofyra}{$\rm{H_2O\,(2_{21}-2_{12})}$}
\newcommand{\htvaofem}{$\rm{H_2O\,(3_{03}-2_{12})}$}
\newcommand{\htvaosex}{$\rm{H_2O\,(3_{12}-2_{21})}$}
\newcommand{\htvaosju}{$\rm{H_2O\,(3_{12}-3_{03})}$}
\newcommand{\cotvaett}{CO\,(2$-$1)}
\newcommand{\cotretva}{CO\,(3$-$2)}
\newcommand{\cofyratre}{CO\,(4$-$3)}
\newcommand{\cofemfyra}{CO\,(5$-$4)}
\newcommand{\cosjusex}{CO\,(7$-$6)}
\newcommand{\cotionio}{CO\,(10$-$9)}
\newcommand{\siotvaett}{SiO\,(2$-$1)}
\newcommand{\siotretva}{SiO\,(3$-$2)}
\newcommand{\siofemfyra}{SiO\,(5$-$4)}
\newcommand{\amin}{$^{\prime}$}                   
\newcommand{\asec}{$^{\prime \prime}$}
\newcommand{\adeg}{$^{\circ}$}

\newcommand{\atwozero}{$\alpha_{2000}$}
\newcommand{\dtwozero}{$\delta_{2000}$}

\newcommand{\radot}[4]{\mbox{#1$^{\rm h}$#2$^{\rm m}$#3$\stackrel{^{\rm
s}}{_{\bf\cdot}}$#4}}
\newcommand{\decdot}[3]{\mbox{#1$^{\circ}$#2$^{\prime}$#3$^{\prime \prime}$}}
\newcommand{\lsun}{$L_{\odot}$}                          
\newcommand{\msun}{$M_{\odot}$}

\newcommand{\mdot}{{\it \.{M}}}
\newcommand{\msunyr}{$M_{\odot} \, {\rm yr}^{-1}$}
\newcommand{\mearth}{$M_{\oplus}$}

\newcommand{\swas}{SWAS} 
\newcommand{\odin}{Odin}
\newcommand{\herschel}{Herschel}
\newcommand{\hifi}{HIFI}
\newcommand{\apex}{APEX}
\newcommand{\sest}{SEST}
\newcommand{\iso}{ISO}

\newcommand{\spitzer}{\textit{Spitzer}}

\newcommand{\mrthank}{${\hspace{5.1mm}}$}

\titlerunning{Herschel observations of the Herbig-Haro objects HH\,52-54}
\authorrunning{P. Bjerkeli et al.}

\begin{document}
\title{Herschel observations of the Herbig-Haro objects HH\,52-54
  \thanks{Herschel is an ESA space observatory with science instruments provided
by European-led Principal Investigator consortia and with important
participation from NASA. \vspace{1.5mm} \newline Complementary
    observations were made with: \vspace{1.5mm} \newline \mrthank
    \odin\ is a Swedish-led satellite project funded jointly by the
    Swedish National Space Board (SNSB), the Canadian Space Agency
    (CSA), the National Technology Agency of Finland (Tekes) and
    Centre National d'Etude Spatiale (CNES). \vspace{1.5mm} \newline
    \mrthank The Swedish ESO Submillimetre Telescope (SEST) located at
    La Silla, Chile was funded by the Swedish Research Council (VR)
    and the European Southern Observatory. It was decommissioned in
    2003. \vspace{1.5mm} \newline \mrthank The Atacama Pathfinder
    EXperiment (APEX) is a collaboration between the
    Max-Planck-Institut f\"{u}r Radioastronomie, the European Southern
    Observatory and the Onsala Space Observatory.}}
\author{P. Bjerkeli\inst{1}, R. Liseau\inst{1}, B. Nisini\inst{2}, M. Tafalla\inst{3}, M. Benedettini\inst{4}, P. Bergman\inst{1}, O. Dionatos\inst{5}, T. Giannini\inst{2}, \\G. Herczeg\inst{6}, K. Justtanont\inst{1}, B. Larsson\inst{7}, C. M$^{\rm{c}}$Coey\inst{8}, M. Olberg\inst{1} and A.O.H Olofsson\inst{1}
                  }
\institute{Department of Earth and Space Sciences, Chalmers University of Technology, Onsala Space Observatory, 439 92 Onsala, Sweden 
\and INAF - Osservatorio Astronomico di Roma, Via di Frascati 33, 00040 Monte Porzio Catone, Italy
\and Observatorio Astron\'{o}mico Nacional (IGN), Calle Alfonso XII,3. 28014, Madrid, Spain
\and INAF - Osservatorio Astrofisico di Arcetri, Largo E. Fermi 5, 50125 Firenze, Italy
\and Centre for Star and Planet Formation, Natural History Museum of Denmark, University of Copenhagen, \O ster Voldgade 5-7, 1350 Copenhagen, Denmark
\and Max Planck Institut for Extraterestrische Physik, Garching, Germany
\and Department of Astronomy, Stockholm University, AlbaNova, 106 91 Stockholm, Sweden
\and University of Waterloo, Department of Physics and Astronomy, Waterloo, Ontario, Canada 
}
   \date{Accepted June 14, 2011}

 \abstract
 { The emission from Herbig-Haro objects and supersonic molecular
   outflows is understood as cooling radiation behind shocks,
   initiated by a (proto-)stellar wind or jet. Within a given object,
   one often observes the occurrence of both dissociative (J-type) and
   non-dissociative (C-type) shocks, owing to the collective effects
   of internally varying shock
   velocities.} 
   { We are aiming at the observational estimation of the relative
     contribution to the cooling by CO and \htvao, as this provides
     decisive information for the understanding of the oxygen
     chemistry behind interstellar shock waves. }
   { The high sensitivity of \hifi, in combination with its high
     spectral resolution capability, allows us to trace the \htvao\
     outflow wings at unprecedented signal-to-noise. From the
     observation of spectrally resolved \htvao\ and CO lines in the
     HH52-54 system, both from space and from ground, we arrive at the
     spatial and velocity distribution of the molecular outflow
     gas. Solving the statistical equilibrium and non-LTE radiative
     transfer equations provides us with estimates of the physical
     parameters of this gas, including the cooling rate ratios of the
     species. The radiative transfer is based on an Accelerated Lambda
     Iteration code, where we use the fact that variable shock
     strengths, distributed along the front, are naturally implied by
     a curved surface.}
   { Based on observations of CO and \htvao\ spectral lines, we
     conclude that the emission is confined to the HH54 region. The
     quantitative analysis of our observations favours a ratio of the
     CO-to-\htvao-cooling-rate $\gg 1$. Formally, we derive the
     ratio $\Lambda({\rm CO})/\Lambda$(\ohtvao) $ =10$, which is
     in good agreement with earlier determination of 7 based on
     ISO-LWS observations. From the best-fit model to the CO
     emission, we arrive at an \htvao\ abundance close to $1 \times
     10^{-5}$. The line profiles exhibit two components, one of
     which is triangular and another, which is a superposed, additional
     feature.
     This additional feature likely originates from a region smaller
     than the beam where the \orthowater\ abundance is smaller than in the quiescent gas.}
   { Comparison with recent shock models indicate that a planar shock
     can not easily explain the observed line strengths and triangular
     line profiles. We conclude that the geometry can play an important
     role. Although abundances support a scenario where J-type shocks are present, higher cooling rate ratios than predicted by these type of shocks are derived. 
     }


   \maketitle

\section{Introduction}

Outflows have many times been discovered through observations of
Herbig-Haro objects \citep[see
e.g.][]{Herbig:1950kx,Herbig:1951vn,Haro:1952fk,Haro:1953uq}, tracing
the gas at highest velocity. \cite{Liseau:1986uq} showed that
HH-objects and outflows are physically associated, implying that they
likely have the same exciting source.

Water is one of the coolants that is most sensitive to different type
of shock chemistry \citep[e.g.][]{Bergin:1998lr}. Depending on the
ionisation fraction, magnetic field strength and velocity of the
shock, water abundances can be elevated to different levels
\citep{Hollenbach:1989ly}. In J-type shocks (Jump shocks), where the
magnetosonic speed is lower than the propagation of the pressure
increase, the involved energies generally dissociate \htva\ and all molecules
with lower binding energies. As such the water abundance is generally
low in J-type shocks, although it may reform in the post-shock cooling
region once pre-shock densities are sufficiently high.
In C-type shocks however, the pre-shock gas is partially heated due to
traversing magnetic waves from the post-shock gas,
and molecules can survive the passage of the shock
\citep{Draine:1980ys} where both the magnetic field and the gas
density is compressed. In this type of shock, the activation barrier
for neutral-neutral reactions between molecular hydrogen and oxygen is
reached, and the water abundance is expected to become enhanced. This
can be both due to the effect of sputtering from dust grains
\citep{Kaufman:1996qy} and due to high temperature chemistry occurring
\begin{figure}[]
  \hfill
  \begin{minipage}[]{0.45\textwidth}
      \hspace{-1.0cm}
      \includegraphics[width = 8.7cm]{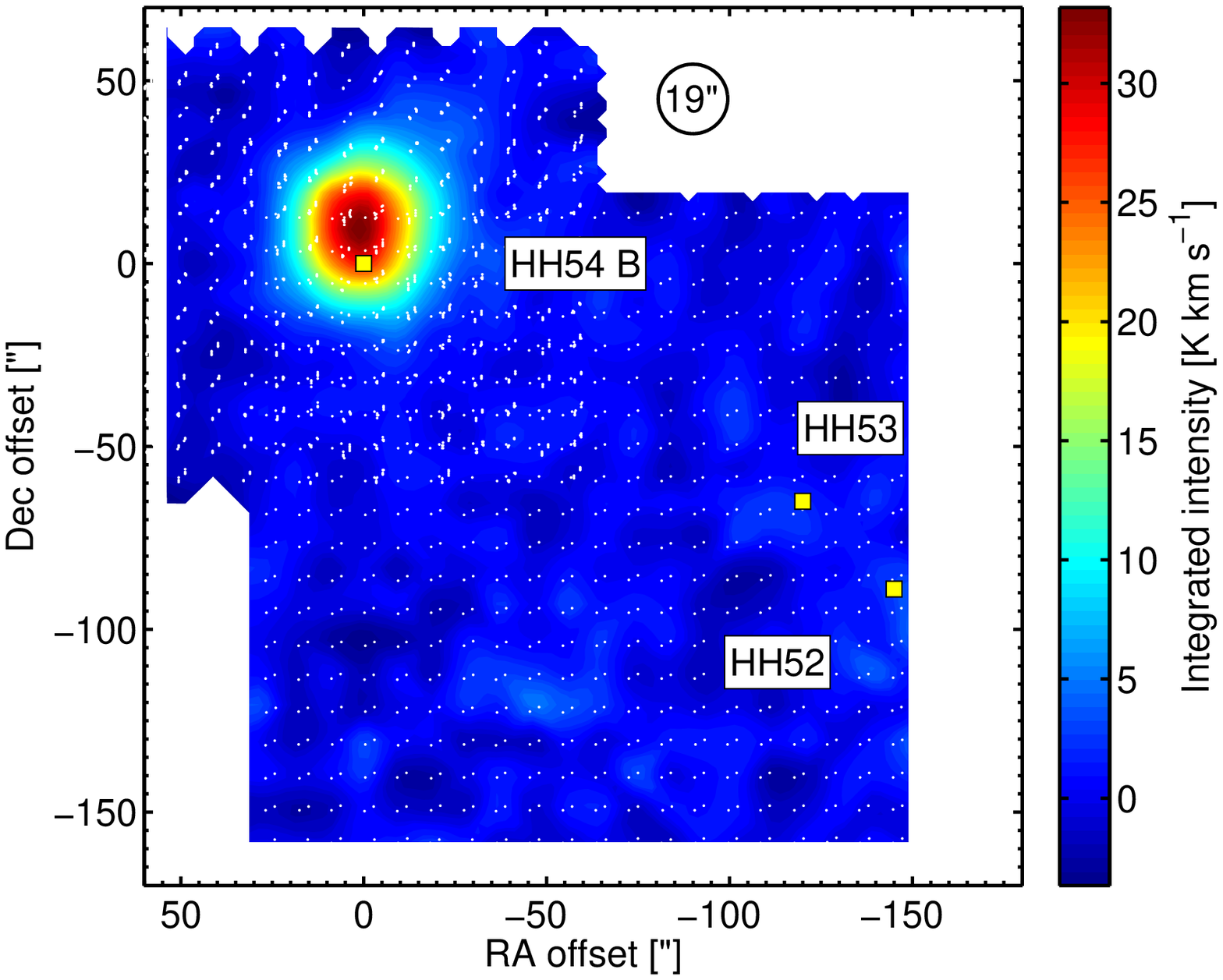}
  \end{minipage}
  \hfill
  \begin{minipage}[]{0.45\textwidth}
\hspace{-0.3cm}
      \includegraphics[width = 7.3cm]{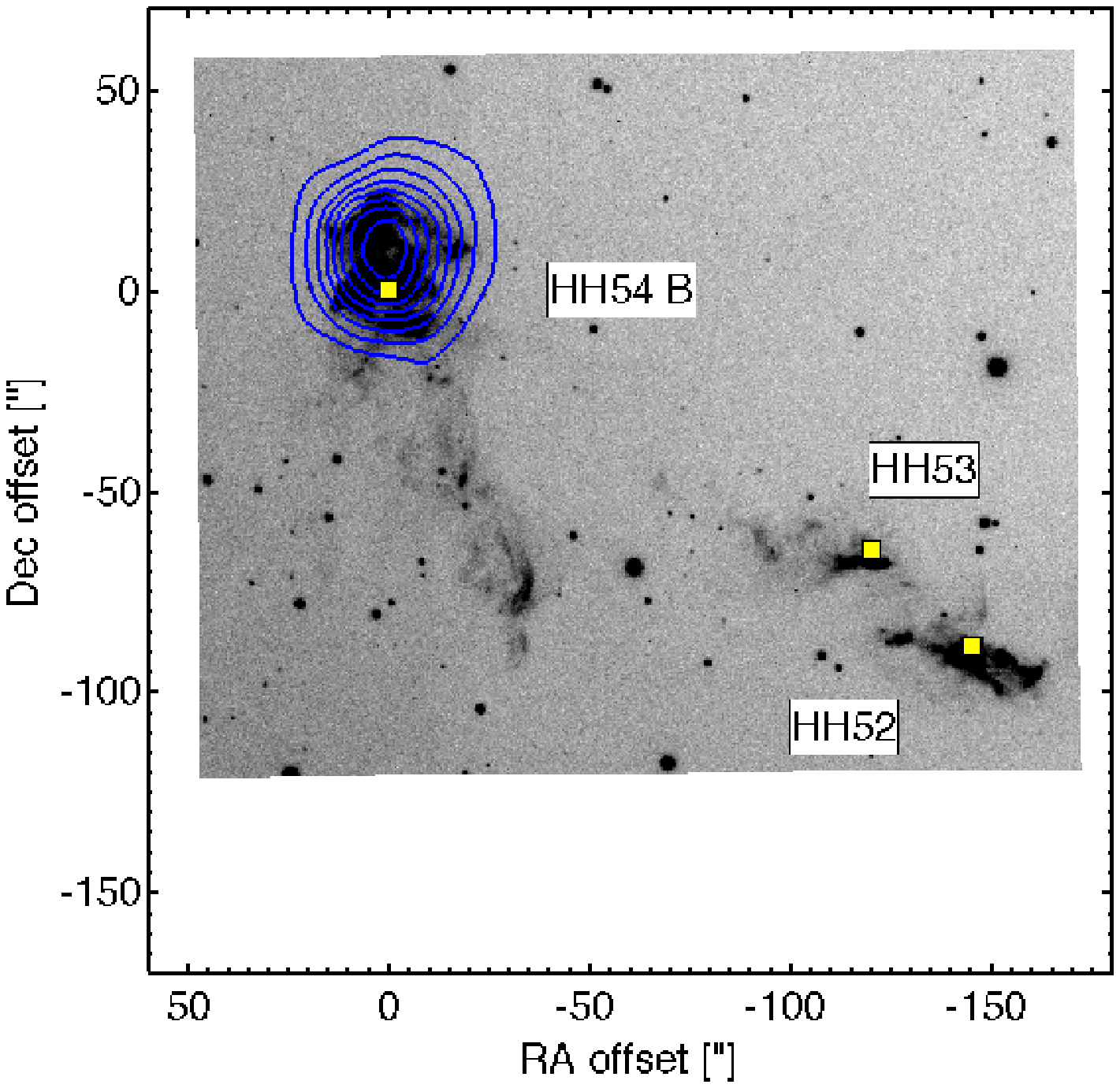}
  \end{minipage}
  \hfill
  \begin{minipage}[]{0.473\textwidth}
      \includegraphics[width = 9.2cm]{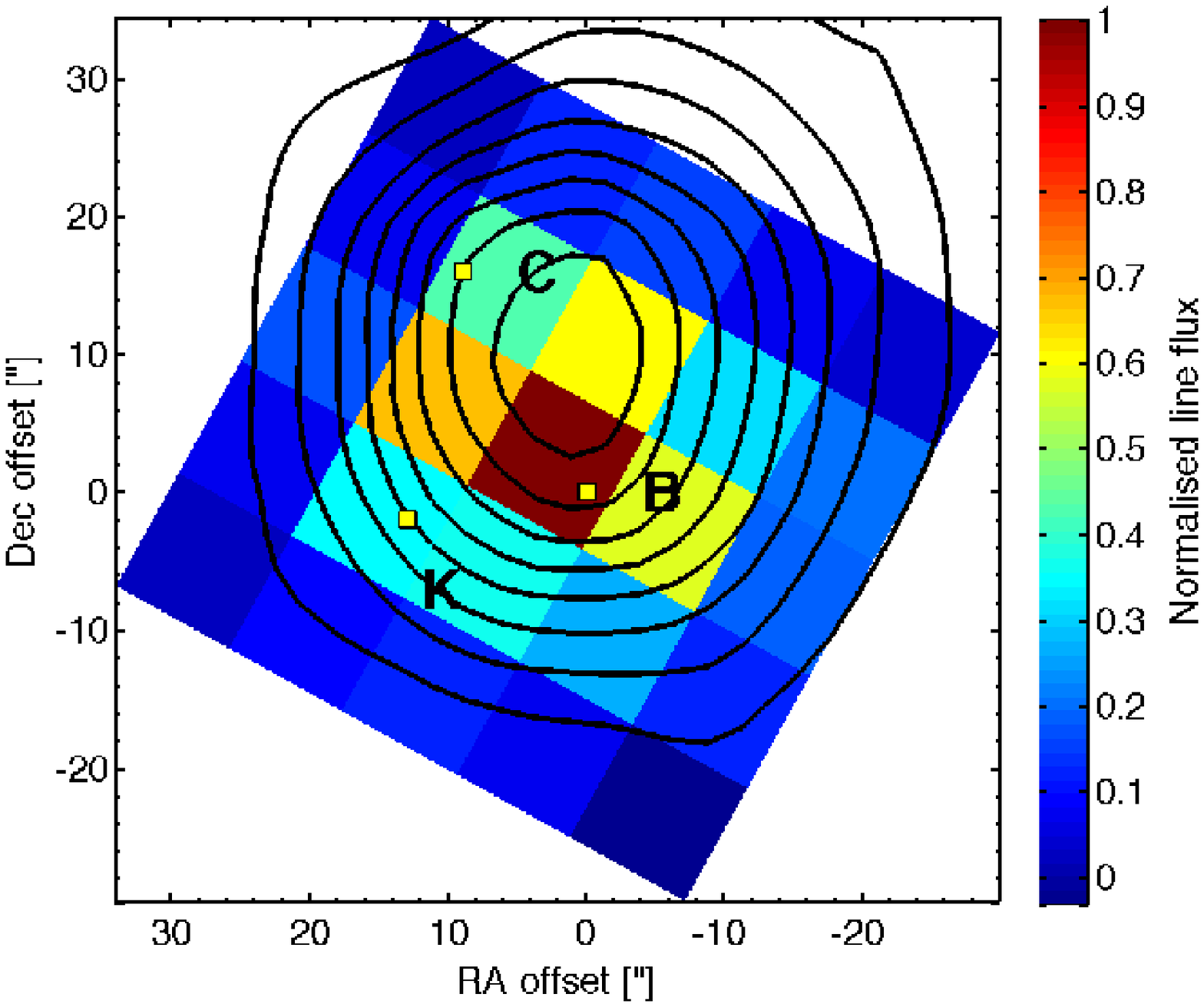}
  \end{minipage}
  \hfill
  \caption{\textit{Upper panel}: False colour map of the \cotionio\
    integrated intensity in the blue line wing (from $+$2.4~\kmpers\
    to $-$30~\kmpers). The positions of HH\,54 B, HH\,53 and HH\,52
   are indicated
    with yellow squares. The readout positions for the on-the-fly and
    raster maps are indicated with white dots. 
    \newline \textit{Middle panel}: \cotionio\ map of the
    integrated intensity obtained with \hifi\ in the blue line wing
    overlaid on an \halpha\ image from
    \citet{Caratti-o-Garatti:2009fk}.  Contours are from 5.5 to 33.6
    K~\kmpers\ in steps of 3.5~K~\kmpers. \newline \textit{Lower
      panel}: A zoom of the \cotionio\ integrated intensity overlaid
    on the \htvaotva\ emission obtained with PACS. The positions of
    HH\,54 B, C and K as indicated in \citet{Giannini:2006lr} are
    indicated with yellow squares.}
  \label{figure:0}
\end{figure}
in the shocked region \citep{Bergin:1998lr}. After the shock passage,
the enhanced water abundance persists for a long time in the
post-shock gas. 

The atmosphere of the Earth is opaque 
at the wavelength of the lowest rotational transitions of water as
well as most other higher excited transitions.
Thus, it is not until recent years, with the use of space based
observatories, these transitions have been observed
successfully. Previous missions such as \swas\ \citep{Melnick:2000fk}
and \odin\ \citep{Nordh:2003qy} have put constraints on the water
abundance and the dynamics of molecular outflows
\citep{Franklin:2008fk,Bjerkeli:2009kx}. None of these missions,
however, provided the spatial and spectral resolution that is
available with the \herschel\ Space
Observatory~\citep{Pilbratt:2010kx}.

HH\,54 is a Herbig-Haro object located in the Chamaeleon~II cloud at a
distance of approximately 180 pc
\citep{Whittet:1997kx}. The visible objects
in HH\,54 are moving at high Doppler speed out of the cloud toward the
observer, with velocities of the order 10~-~100~\kmpers\
\citep{Caratti-o-Garatti:2009fk}. The objects show a clumpy appearance
due to either Rayleigh-Taylor instabilities in the flow or due to
variability in the jet itself. Another possibility is patchy overlying
dust extinction. The source of the jet is not well constrained. This
is further discussed in appendix~\ref{section:source}, where we also
present quantitative arguments for identifying IRAS 12553-7651 (ISO-Cha
II 28) as the HH54 jet-driving source and its associated blueshifted
CO outflow. No redshifted emission is observed as what would be
expected from a bipolar jet (see Section~\ref{section:results}). On the
other hand, the extinction is relatively low, something that might
allow the redshifted gas to flow out essentially unhindered into the
low density material at the rear side of the cloud. It can however not
be ruled out that the outflow itself is one-sided and
asymmetric. Recent simulations show that rapidly rotating stars with
complex magnetic fields can be responsible for such type of flows
\citep{Lovelace:2010fk}.

HH\,54 has previously been observed in various lines of CO, SiO and
\htvao\ using \odin\ and SEST
\citep[][R.~Liseau,~unpublished]{Bjerkeli:2009kx}. Several \htvao\
lines and high-J CO lines were also observed with ISO-LWS and
published in \citet{Liseau:1996fk} and \citet{Nisini:1996fk}. During
the Performance Verification Phase of the Heterodyne Instrument for
the Far-Infrared (HIFI) instrument, aboard the \herschel\ Space
Observatory, the \cotionio\ transition was observed. As part of the
WISH keyprogram \citep{van-Dishoeck:2011lr}, also the \htvaoett\ and \htvaotva\
transitions were observed using HIFI and the Photodetector Array
Camera and Spectrometer (PACS) respectively. We note that the
wavelength region, covering the \htvaotva\ transition, also can be
observed with HIFI.

In this paper, we present observations, both from space and ground, of
HH\,52-54 in spectral lines of CO and \htvao. We choose to observe
this region based on the fact that HH54 is free from contamination
from other objects, spatially confined and resolved in the infrared
regime with the instruments used. Using results from observations
carried out with \sest, \apex, \odin\ and \herschel, we aim at
improving our understanding of interstellar shock waves. The observing
modes and the instruments that have been used are described in
Section~\ref{section:observations} while the details of the \hifi\
data reduction can be found in Appendix~\ref{section:reduction}. The
basic observational results are summarised in
Section~\ref{section:results} whereas the interpretations are
discussed in
Section~\ref{section:discussion}. 
\begin{table*}[t]
  \begin{center}
    \caption{Molecular line observations carried out with \odin, \sest, \apex\ and \herschel.}
      \label{table:log}
      \renewcommand{\footnoterule}{} 
      \begin{tabular}{l l r r r c c r}
        \hline
        \hline
        \noalign{\smallskip}
        Telescope & Molecule & Frequency & $E_u/k_{\rm{B}}$& HPBW & $\eta_{\rm{mb}}$& Date & $t_{\textrm{\tiny int}}$\\
        & & (GHz)& (K) & (\asec)& & (YYMMDD) & (hr)  \\
        \noalign{\smallskip}
        \hline
        \noalign{\smallskip}
        \noalign{\smallskip}
        \sest & \cotvaett &230.538&16.6&23&0.50& 970811 - 980806 & 3 \\   
        \sest & \cotretva\ &345.796&33.2&15&0.25& 970811 - 980806 & 3 \\ 
        \apex & \cofyratre\ &461.041&55.3&14&0.60& 070918 & 0.5 \\
        \odin\ & \cofemfyra\ &576.268& 83.0 &118&0.90& 050502 - 050620 & 4\\
        \apex & \cosjusex\ &806.652&155.9&8&0.43& 070918 & 0.8\\
        \herschel-HIFI & \cotionio\ &1151.985&304.2&19&0.66& 090726 - 100221 & 9\\
        \odin\ & \htvaoett\ &556.936&42.4 &126&0.90& 090609 - 100420 & 12\\
        \herschel-HIFI & \htvaoett\ &556.936& 42.4 & 39 & 0.76 & 100729 & 0.05 \\
        \herschel-PACS & \htvaotva\ &1669.905&79.5&13&N/A& 090226 & 0.1\\
        \noalign{\smallskip}
        \noalign{\smallskip}
        \hline
      \end{tabular}
  \end{center}
\end{table*}

\section{Observations}
\label{section:observations}
The observations described in this paper were obtained between 1997
and 2010 with several different facilities. A summary of the observations is presented in
Table~\ref{table:log} while the line intensities are listed in
Table~\ref{table:results}.
\subsection{\herschel}
\subsubsection{HIFI}
During the Performance Verification Phase of HIFI
\citep{de-Graauw:2010uq}, the \cotionio\ data were obtained on 26-27
July 2009 and 21 February 2010. The \htvao\ data were obtained 29 July
2010. The 3.5~m Cassegrain
telescope has a Full Width Half Maximum (FWHM) of 38\asec\ at 557~GHz,
19\asec\ at 1152~GHz and 13\asec\ at 1670~GHz,
respectively. The \hifi\ \htvaoett\ spectrum presented in
this paper was obtained in point mode with position switch using band
1 (490 - 630~GHz). The OFF spectrum is obtained by a single
observation of a reference point 10\amin\ away. The \cotionio\ \hifi\ maps
were obtained in two different observing modes using band 5 (1120 -
1250 GHz). In the dual-beam-switch raster mode, an internal chopper
mirror is used to obtain an OFF spectrum 3\amin\ away from the
observed position. In the on-the-fly with position switch mode, the
telescope is scanning the map area back and forth.
The data were calibrated using the \textit{Herschel Interactive
  Processing Environment} (HIPE) version 4.2 and 5.0 for the
\cotionio\ and \htvaoett\ observations, respectively
\citep{Ott:2010mz}. The data reduction of the HIFI maps is described in
detail in Appendix~\ref{section:reduction}. For the \cotionio\
observation, data from one of the spectrometers on-board have been
used. The Wide Band Spectrometer (WBS) is an acousto-optical
spectrometer with a 4~GHz frequency coverage. The channel spacing is
500~kHz (0.1~\kmpers\ at 1152~GHz and 0.3~\kmpers\ at 557~GHz). For
the \htvaoett\ observation, data from the High Resolution Spectrometer
(HRS) have also been used. The HRS is an Auto-Correlator System (ACS)
where the resolution can be varied from 0.125 - 1.00~MHz. For this
observation it was set to 0.24~MHz. Observations from the horizontal
(H) and the vertical (V) polarisations were combined for both
observations. The spectra were converted to a \tmb\ scale using main
beam efficiencies, $\eta_{\mathrm{mb}}(1152 \mathrm{GHz})$ = 0.66 and
$\eta_{\mathrm{mb}}(557 \mathrm{GHz})$ = 0.76 \citep{Olberg:2010kx}.
\subsubsection{PACS}
The PACS spectrograph \citep{Poglitsch:2010uq} is a 5\ggr5 integral
field unit array consisting of $9\farcs4$ square spaxels (spatial
picture elements). The observations were obtained on 26 February 2009
in line scan mode and cover the 178.8--180.5~\um\ region, centered on
the \htvaotva\ line at 179.5~\um. 
The blue channel simultaneously covered the 89.4--90.2~\um\ region,
which is featureless and not discussed further.  Two different nod
positions, located 6\amin\ from the target in opposite directions,
were used to correct for the telescopic background. Data were reduced
with HIPE version 4.0.  The fluxes were normalised to the telescopic
background and subsequently converted to an absolute flux based on
PACS observations of Neptune \citep{Lellouch:2010zr}, with an
approximate uncertainty of \about20 \% at 180~\um. The spatial
resolution at 180~\um\ is nearly diffraction-limited (see Table~\ref{table:log}).
In well-centered observations of point sources, only about 40~\% of
the light in the system falls within the central spaxel.  The
\htvaotva\ line is spectrally unresolved in the $R=1700$ spectra
(\deltav~\about~175~\kmpers). 

\subsection{\odin}
The \odin\ space observatory carries a 1.1~m Gregorian telescope and
was launched into space in 2001
\citep{Nordh:2003qy,Hjalmarson:2003kx}. It is located in a polar orbit
at 600~km altitude. At 557 and 576~GHz, the FWHM is 126\asec\ and 118\asec\
respectively. The
spectra were converted to a \tmb\ scale using a main beam efficiency,
$\eta_{\mathrm{mb}}$~=~0.9, as measured from Jupiter observations
\citep{Hjalmarson:2003kx}.

The \cofemfyra\ observations, suffered from frequency drift
and have, for that reason, been calibrated, using atmospheric spectral
lines, acquired during the time intervals when \odin\ observed through
the Earth's atmosphere \citep{Olberg:2003fj}. Since the first
publication of the \cofemfyra\ data in \citet{Bjerkeli:2009kx}, the
frequency calibration scheme has improved.  Despite
this, the velocity scale for this particular observation has some
uncertainties due to in-orbit variations of the local oscillator unit
frequency. These variations are most likely caused by slight
temperature changes in the spacecraft during each orbit, due to the
fact that the Earth is located very nearby. On a velocity scale these
fluctuations correspond to a \about2 \kmpers\ uncertainty. 

A spectrum, showing a tentative detection of \htvaoett\
at 557~GHz was published in \citet{Bjerkeli:2009kx}. Since
then, however, additional observations toward HH\,54 have
been carried out in June 2009 and April 2010 for a total on-source
time of 12 hours.

The observing mode for both observations was position switching, where
the entire telescope is re-orientated to obtain a reference spectrum
(10\amin\ away in June 2009 and 15\amin\ away in April 2010). The spectrometer used is an acousto-optical
spectrometer (AOS) where the channel spacing is 620~kHz (0.33~\kmpers\
and 0.32~\kmpers\ at 557~GHz and 576~GHz respectively). The data
processing and calibration are described by \citet{Olberg:2003fj}.

\subsection{\sest}
The observations and calibration of the \cotvaett, \cotretva,
\siotvaett, \siotretva\ and \siofemfyra\ data obtained with \sest\
were already described in detail by \citet{Bjerkeli:2009kx}, to which
we refer the interested reader.

\subsection{\apex}
\cofyratre\ and \cosjusex\ data were obtained with the APEX/FLASH
receiver in service mode in July 2006.  
In this project\footnote{ESO project code: 077.C-4005(A)}, both CO
lines were observed simultaneously in a 
grid map around HH54\,B centered on \atwozero~=~\radot{12}{55}{49}{5},
\dtwozero~=~\decdot{$-$76}{56}{23}. The selected reference position
was relatively nearby at (120\asec,$-$120\asec), which resulted in
contaminated line spectra close to the cloud LSR velocity (see
below). The telescope pointing was checked by observing the nearby
nova X Tra (IRAS 15094-6953). The map was spaced by half the
instrument beam for the \cofyratre\ transition which corresponds to
7\asec\ at 450~GHz. The data reduction was performed in CLASS and the spectra were converted
to a \tmb\ scale using the main beam efficiencies,
0.60 and 0.43 for the \cofyratre\ and \cosjusex\ transitions,
respectivelly \citep{Gusten:2006fk}. The channel spacing is 61~kHz
(0.04~\kmpers\ at 461~GHz and 0.02~\kmpers\ at 807~GHz).
\section{Results}
\label{section:results}
The results from our observations are summarised in
Table~\ref{table:results}, where the integrated intensity over the
blue line wing is presented. 
\begin{table}[t]
  \begin{flushleft}
    \caption{Integrated intensities over the line wings and 1 $\sigma$ \newline uncertainties in parentheses. }
    \resizebox{\hsize}{!}{
      \label{table:results}
      \renewcommand{\footnoterule}{} 
      \begin{tabular}{llcrr}
        \noalign{\smallskip}
      \hline
      \hline
      \noalign{\smallskip}
      Line & Source & $\Delta$\vlsr & $ \int\! T_{\rm{mb}}~$\dv& $T_{\rm{mb,rms}}$$^{a}$   \\
      \noalign{\smallskip}
      &  &{($\rm{km~s^{-1}}$)} &({$\rm{K~km~s^{-1}}$}) &  (mK)  \\
      \noalign{\smallskip}
      \hline          
      \noalign{\smallskip}
      \cotvaett &HH\,54 & 0.9 to $-23.3$& 34.6 (0.3) & 224 \\
      \cotretva\ &HH\,54 & 0.9 to $-23.3$& 54.9 (1.9) & 2021 \\
      \cofyratre\ & HH\,54 &2.4 to $-23.3$ & 104.4 (3.9) & 773 \\
      \cofemfyra\ &HH\,54 & 2.4 to $-21.8$ & 9.3 (0.06) & 20 \\
      \cosjusex\ & HH\,54 & 2.4 to $-21.8$ & 128.1 (25.3) & 4741 \\
      \cotionio\ &HH\,52 & - & - & 284 \\
      &HH\,53 & - & - &211 \\
      &HH\,54 & 2.4 to $-21.8$ & 33.1 (0.3) & 146 \\
      \htvaoett$^{b}$ &HH\,54 &2.4 to $-22.8$& 9.2 (0.06) & 20 \\
      \noalign{\smallskip}
      \noalign{\smallskip}
      \hline
    \end{tabular}
}
    Notes to the Table: $^{a}$The velocity bin size when calculating the rms is the same as the channel spacing. $^{b}$This refers to the spectra obtained with \hifi.
\end{flushleft}
\end{table}
\subsection{\herschel}
\label{section:resultsherschel}
\begin{figure}[t]
    \begin{center}
      \rotatebox{270}{\includegraphics[width=6.4cm]{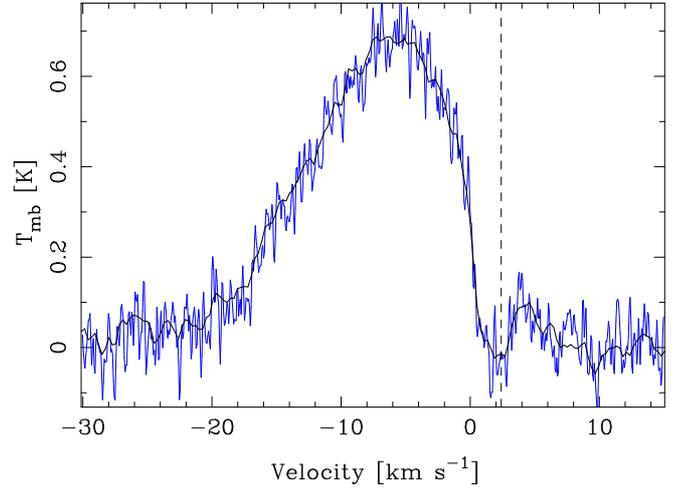}}
    \end{center}
   \caption{
\htvaoett\ spectra obtained with \hifi. The blue
    spectral line is the HRS and the black spectral line is the
    WBS. Both observations were centered on HH\,54 B:
    \atwozero~=~\radot{12}{55}{50}{3},
    \dtwozero~=~\decdot{$-$76}{56}{23}. The dashed line indicates the
    position of the velocity of the cloud.}
  \label{figure:hifiwaterspectrum}
\end{figure}

\cotionio\ is only detected in the HH\,54 region. No emission is
detected toward the region of HH\,52-53 down to an rms level of
\about0.2~K (see upper panel of Fig.~\ref{figure:0}) and these
objects will not be discussed further. From the \cotionio\ emission, 
we estimate the size of the source to \about27\asec. The \htvaoett\
line is clearly detected toward HH\,54 (see Fig.~\ref{figure:hifiwaterspectrum}).
Simultaneously, the NH$_3(1_0-0_0)$ line at 572~GHz was covered in the
upper side band. No emission is detected down to an rms level of
\about20~mK. For the \htvaotva\ line observed with PACS, emission is
detected in most of the spaxels and the angular extent of the source
is no larger than \about28\asec. The peak flux in the central spaxel
is 9 Jy. 
\begin{figure*}[t]
  \resizebox{\hsize}{!}{
    \rotatebox{270}{\includegraphics[width=9cm]{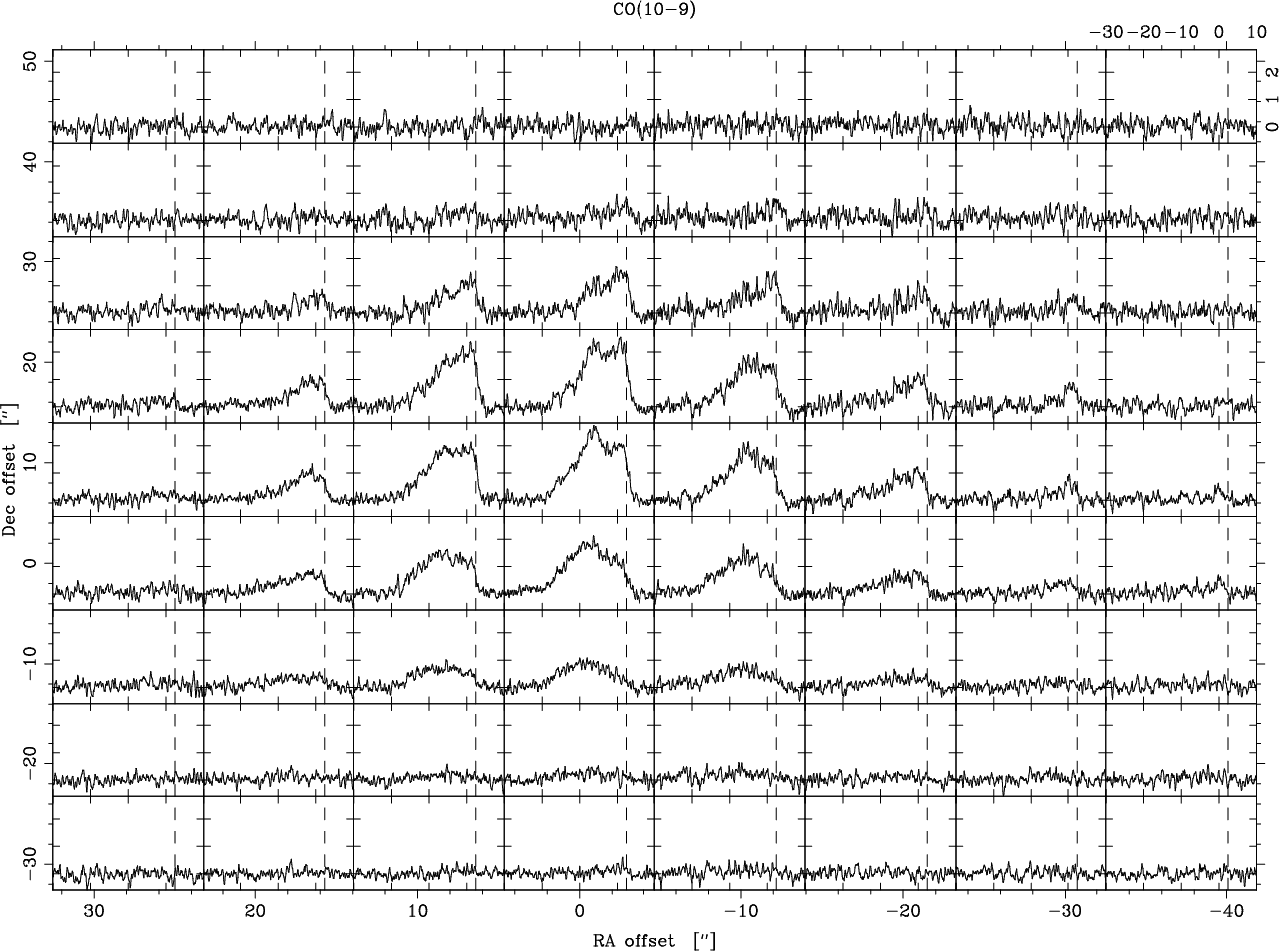}}
  }
  \caption{\cotionio\ map obtained with \hifi. The map shows the
    spectra toward the region close to HH\,54 B and the map has been
    regridded with map spacing equal to $9\farcs3$. Offsets are with
    respect to HH\,54 B: \atwozero~=~\radot{12}{55}{50}{3},
    \dtwozero~=~\decdot{$-$76}{56}{23}. The velocity scale (\vlsr) and
    intensity scale ($T_{\rm{mb}}$) is indicated in the upper right
    corner of the map, \vlsr~=~+2.4~\kmpers\ with a dashed
    line.}
  \label{figure:CO10-9map}
\end{figure*}
In the lower panel of Figure~\ref{figure:0}, the \cotionio\ integrated
intensity contours are shown overlaid on the \htvaotva\ normalised
flux in each spaxel. In this figure, each spaxel is presented on a
square grid. In reality however, there is a small misalignment between
each spaxel \citep[see][Their
Fig~10]{Poglitsch:2010uq}. An offset of
$\sim$9\asec\ between the \cotionio\ and the \htvaotva\ emission peak
is also observed, where the peak of the CO emission is located in
between the B and C clumps as identified by
\cite{Sandell:1987vn}. This offset might be real given a
pointing accuracy of a few arcseconds for \herschel. Noteworthy is
that the different clumps in the region show detectable proper motion
over a time scale of a few years
\citep[e.g.][]{Schwartz:1984kx,Caratti-o-Garatti:2006lr,Caratti-o-Garatti:2009fk}. However,
the CO and \htvao\ observations with \herschel\ were obtained over a
time span of only one year and it is therefore unlikely that proper
motion is the cause of the observed offset. The \htvaoett\
spectra obtained with \hifi\ are presented in
Fig.~\ref{figure:hifiwaterspectrum}. The line is self absorbed by the
foreground cloud at \vlsr~=~+2.4~\kmpers. 
\subsection{\odin}
The \odin\ \htvaoett\ observations carried out on 9 June 2009 and
20 April 2010 confirmed the previously published detection with an improved
signal to noise. It is this dataset that is used for the
comparison with \hifi\ data in the present paper. 

\subsection{\sest}
\begin{figure*}[ht]
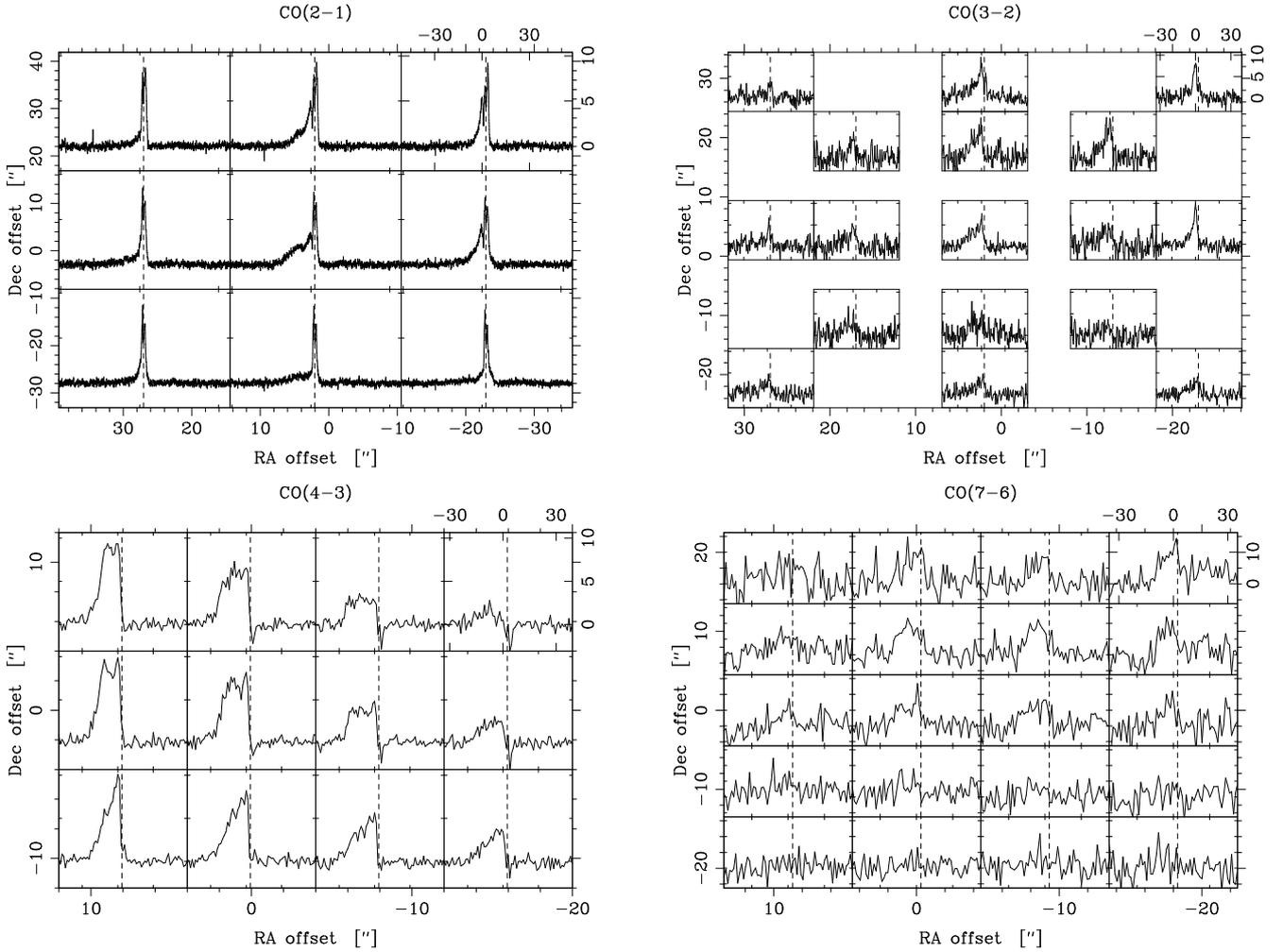

\begin{minipage}[b]{0.5\linewidth}
\centering
\rotatebox{270}{\includegraphics[scale=0.35]{CO2-1map.ps}}
\end{minipage}
\hspace{0cm}
\begin{minipage}[b]{0.5\linewidth}
\centering
\rotatebox{270}{\includegraphics[scale=0.35]{CO3-2map.ps}}
\end{minipage}
\begin{minipage}[b]{0.5\linewidth}
\centering
\vspace{0.2cm}
\rotatebox{270}{\includegraphics[scale=0.35]{CO4-3map.ps}}
\end{minipage}
\begin{minipage}[b]{0.5\linewidth}
\centering
\rotatebox{270}{\includegraphics[scale=0.35]{CO7-6map.ps}}
\end{minipage}
\caption{Same as Figure~\ref{figure:CO10-9map}, but for the \cotvaett, \cotretva, \cofyratre\ and \cosjusex\ maps obtained with \sest\ and \apex. The \apex\ maps have been regridded with a map spacing equal to 8\asec\ for \cofyratre\ and 9\asec\ for \cosjusex. Note that the \cofyratre\ spectra may be contaminated with OFF beam contribution.}
\label{fig:COmaps}
\end{figure*}

The \cotvaett\ and \cotretva\ maps were centered with a slight offset
with respect to HH\,54 B, viz $4\farcs8$. The spacing in \cotvaett\
was 25\asec while the spacing between the observations in \cotretva\
were of the order 15\asec, i.e one full beam width. The number of
positions observed in the \cotvaett\ map and the quality of the
baselines in the \cotretva\ does not allow us to put constraints on
the source size.  SiO was not detected, when averaging all spectra
together, down to an rms level of 10 mK for \siotvaett, 15 mK for
\siotretva\ and 7 mK for \siofemfyra. The \sest\
data are presented in Fig.~\ref{fig:COmaps}.
\subsection{\apex} 
The quality of the FLASH data is not fully satisfactory, as they
suffer from off-beam contamination near the line centre.  However, as
we here are mainly focussing on the line wings, this should affect our
conclusions only little, if at all. In these CO line maps, a
bump feature (See Sec.~\ref{section:obsline}) in the line profile was detected in some positions (at
\vlsr~=~$-$7~\kmpers, see below). 
A reliable source size from these maps can however not be determined.
In both maps, the feature is detected
in a velocity range spanning over \deltav~$\simeq$~7~\kmpers\ (see
Fig.~\ref{fig:COmaps}) .
\section{Discussion}
\label{section:discussion}
\begin{table}[t]  
\flushleft
\caption{Parameters used in the CO model}          
\label{table:1}     
\begin{tabular}{l l}          
  \hline\hline                        
  \noalign{\smallskip}
  \noalign{\smallskip}
  \textit{Parameters held constant} & \\
  \noalign{\smallskip}
 Distance to source & $D\mathrm{_{source}}$ = 180 pc\\
  CO abundance & \xco~=~8\texpo{-5}  \\
  LSR velocity & \vlsr~=~2.4~\kmpers \\
  Velocity profile &$\upsilon(r)~=~20\,r / R_{\rm max}$~\kmpers\\ 
  Shell thickness &$\Delta r / R_{\rm max}$ = 0.12  \\
 Source size &$\mathrm{\theta_{source}} = 30$\asec\\
  Microturbulence & $v\mathrm{_{turb} = 1.5\,\,km\,\,s^{-1}}$  \\
  Gas to dust mass ratio &$M\mathrm{_{gas}/}M\mathrm{_{dust}}$ = 100  \\
  Emissivity parameter & $\kappa_{\mathrm{250~\mu m}}$ = 25 $\mathrm{cm^{2}\,g^{-1}}$\\ 
   Dust frequency dependence & $\beta$ = 1\\
  \noalign{\smallskip}
   \noalign{\smallskip}
   \textit{Free parameters}& \\
   \noalign{\smallskip} 
\htva\ density &\nhtva~=~1\texpo{4} - 1\texpo{8}~\cmthree  \\
Kinetic temperature &\tkin~=~10 - 330~K\\
  \noalign{\smallskip}
  \noalign{\smallskip}
  \hline                                             
  
\end{tabular}

\end{table}
\subsection{Observed line profiles}
\label{section:obsline}
\begin{figure*}[]
\begin{minipage}[b]{0.5\linewidth}
\centering
\rotatebox{0}{\includegraphics[scale=0.5]{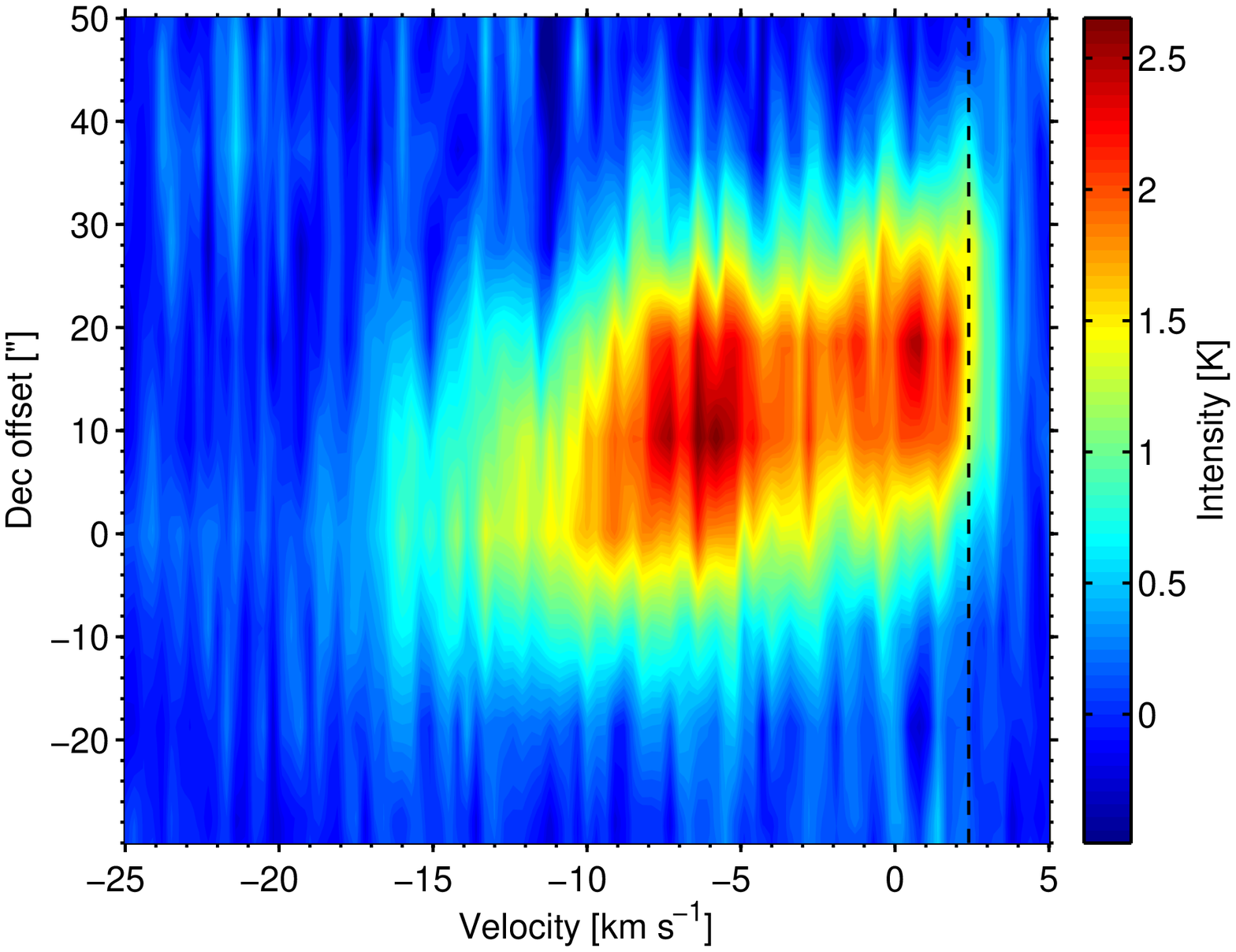}}
\end{minipage}
\hspace{0cm}
\begin{minipage}[b]{0.5\linewidth}
\centering
\rotatebox{0}{\includegraphics[scale=0.5]{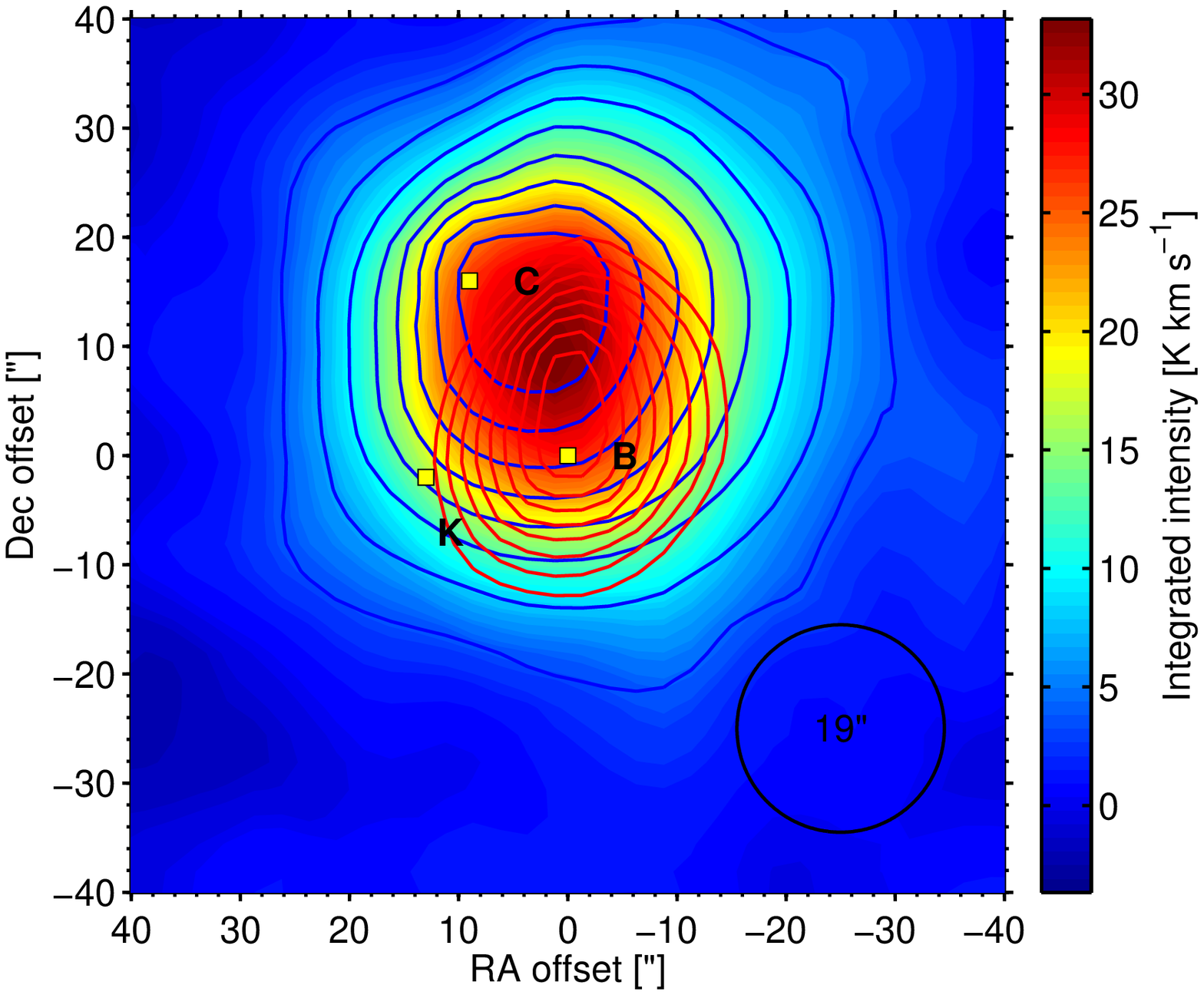}}
\end{minipage}
\caption{\textit{Left panel}: Position (in DEC) velocity diagram of
  the \hifi\ \cotionio\ data showing the observed intensity. The
  bump-like feature can be seen at \vlsr~=~$-$7~\kmpers. The offset in
  RA is 0 \asec. \vlsr~=$+$2.4~\kmpers\ is indicated with a dashed
  line. \newline \textit{Right panel}: Blue contours are the
  integrated intensity when the bump-like feature is subtracted from
  the spectra. Contours are from 3.6 to 28.7~K~\kmpers\ in steps of
  3.1 K~\kmpers. Red contours show the integrated intensity for the
  bump. Contours are from 2.6 to 7.5~K~\kmpers\ in steps of
  0.6~K~\kmpers. Underlying colors show the total integrated intensity
  over the observed line. The positions of HH\,54 B, C and K are
  indicated with yellow squares.}
\label{fig:posvel}
\end{figure*}

Common to the observed transitions in CO and \htvao\ is that only
blue-shifted emission is detected. For all transitions, the maximum
detected velocity in the line wing is of the order of $-20$~\kmpers. A
bump-like feature at \vlsr~$\eqsim -7$~\kmpers\ is also clearly
visible in the observed \cotionio\ and \cotvaett\ spectra and possibly
also in the \cotretva, \cofemfyra\ and \htvaoett\ data. In the
\cotionio\ map, this feature is more prominent in some positions than
others, and most likely it is spatially unresolved to \herschel\ (see
Fig~\ref{figure:CO10-9map}). The \vlsr~$\simeq -7$~\kmpers\ component
is also clearly visible in the \cofyratre\ and \cosjusex\ spectra
observed with \apex\ (see Figure~\ref{fig:COmaps}) where the beam
sizes are 13\asec\ and 8\asec, respectively. Also in these maps, this
component seems unresolved, hence it likely originates from a region
with an angular extent that is smaller than the telescope
beams. In the \cofemfyra\ data, where the beam size is 118\asec, the bump-like feature is barely visible.  This is expected if the beam filling factor is small. A position-velocity diagram of the \cotionio\ transition shows a
trend of higher velocities being detected at lower declination and
closer to the position of HH\,54 B (see
Figs.~\ref{figure:CO10-9map}~and~\ref{fig:posvel}), i.e where the
\htvaotva\ emission peaks (see Fig.~\ref{figure:0}). This is also
clearly visible in the right panel of Fig.~\ref{fig:posvel} where the
integrated intensity of the bump is presented together with the
integrated intensity for the underlying triangular profile. The
intensity maximum of the bump seems offset by $\sim$10\asec\ to
the south from the peak of the bulk outflow emission. The uncertainty
attributed to this offset due to the baseline subtraction is of the
order $\sim$5\asec.
Assuming that the apex of the shock is located close to the HH54\,B
position, one would also expect the highest velocities in this
region.

\subsection{Interpretation of the emission line data}
To compute the line profiles for the observed emission, we use an Accelerated Lambda Iteration (ALI) code (See App.~\ref{section:radtrans}). The ALI code we use is a non-LTE, one dimensional code, assuming spherical geometry where several subshells are used. The number of cells and angles used in the ray tracing can also be arbitrarily chosen. In this work, a curved geometry is compared with a plane parallel slab to interpret the observed spectral lines.

The formation of the observed spectral lines occurs most likely
in shocked gas. Using results from detailed models of C-shocks,
\citet{Neufeld:2006fk} presented estimates of the excitation
conditions for HH\,54. From the analysis of \molh-rotation diagrams,
\citet{Neufeld:2006fk} determined the presence of two different
temperature regimes, viz. at 400~K and at \powten{3}~K, respectively.

We used their analytical expressions for the column density, 
\begin{equation}
N(\mathrm{H_2}) = 7.9 \times 10^{20} \left[\frac{n(\mathrm{H_2})}{10^5}\right]^{0.5} \left(\frac{T_{\mathrm{gas}}}{1000}\right)^{-0.555}~\mathrm{cm^{-2}},
\label{eq:NH2}
\end{equation}
and for the velocity gradient,
\begin{equation}
\frac{d\upsilon}{dz} = 1.6 \times 10^4 \left[\frac{n(\mathrm{H_2})}{10^5}\right]^{0.5} \left(\frac{T_{\mathrm{gas}}}{1000}\right)^{1.30}~\mathrm{km~s^{-1}~pc^{-1}}.
\label{eq:dvdz}
\end{equation}
to compute the emission in the CO and \htvao\
lines.
Here $n(\rm H_2)$
is the pre-shock density and the other quantities have their usual
meaning. Following \citet{Neufeld:2006fk} we take the compression
factor of 1.5, which was used by them for \htva, but we assume that 
holds for CO as well. In addition, they assume that the fractional beam
filling of the two-temperature components is given by the ratio of the
column density derived from the rotation diagram and that given by
Eq.~\ref{eq:NH2}.

For these two temperatures, and for pre-shock
densitites\footnote{These were the pre-shock densities used by
  \citet{Neufeld:2006fk} to compute the \htva\ abundance.} of
\powten{4}~\cmthree\ and \powten{5}~\cmthree, we compute from,
Eqs.~(\ref{eq:NH2}) and (\ref{eq:dvdz}) in slab geometry, the emission
in CO and \htvao\ lines, including the shapes of the lines. Other
parameters (such as CO abundance, source size, microturbulence etc.)
are given in Table~\ref{table:1}. For the \htvao\ abundance we assume
\xhtvao~=~1.0\texpo{-5}, consistent with the upper limit of
$<$~2\texpo{-5} determined by \citet{Neufeld:2006fk}. 

\begin{figure}[t]
\begin{flushleft}
   \includegraphics[width=8cm]{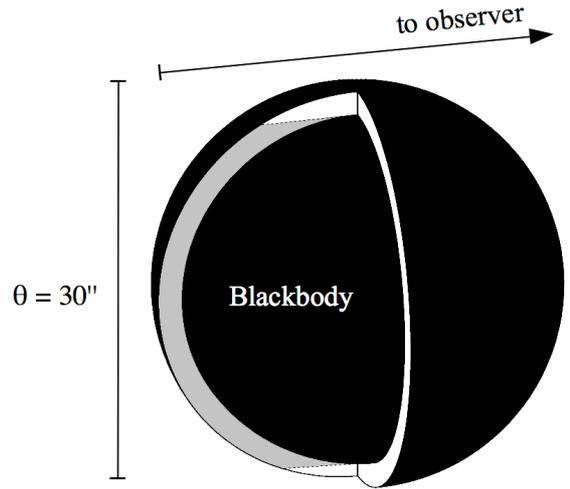}
     \caption{A cut through the shells of the spherical model described in the text. Essentially all radiation originating from the rear side of the sphere (grey) is blocked out by the central blackbody source.}
     \label{figure:modelsphere}
\end{flushleft}
\end{figure}

\begin{figure*}[t]
\centering
   \includegraphics[width=14cm]{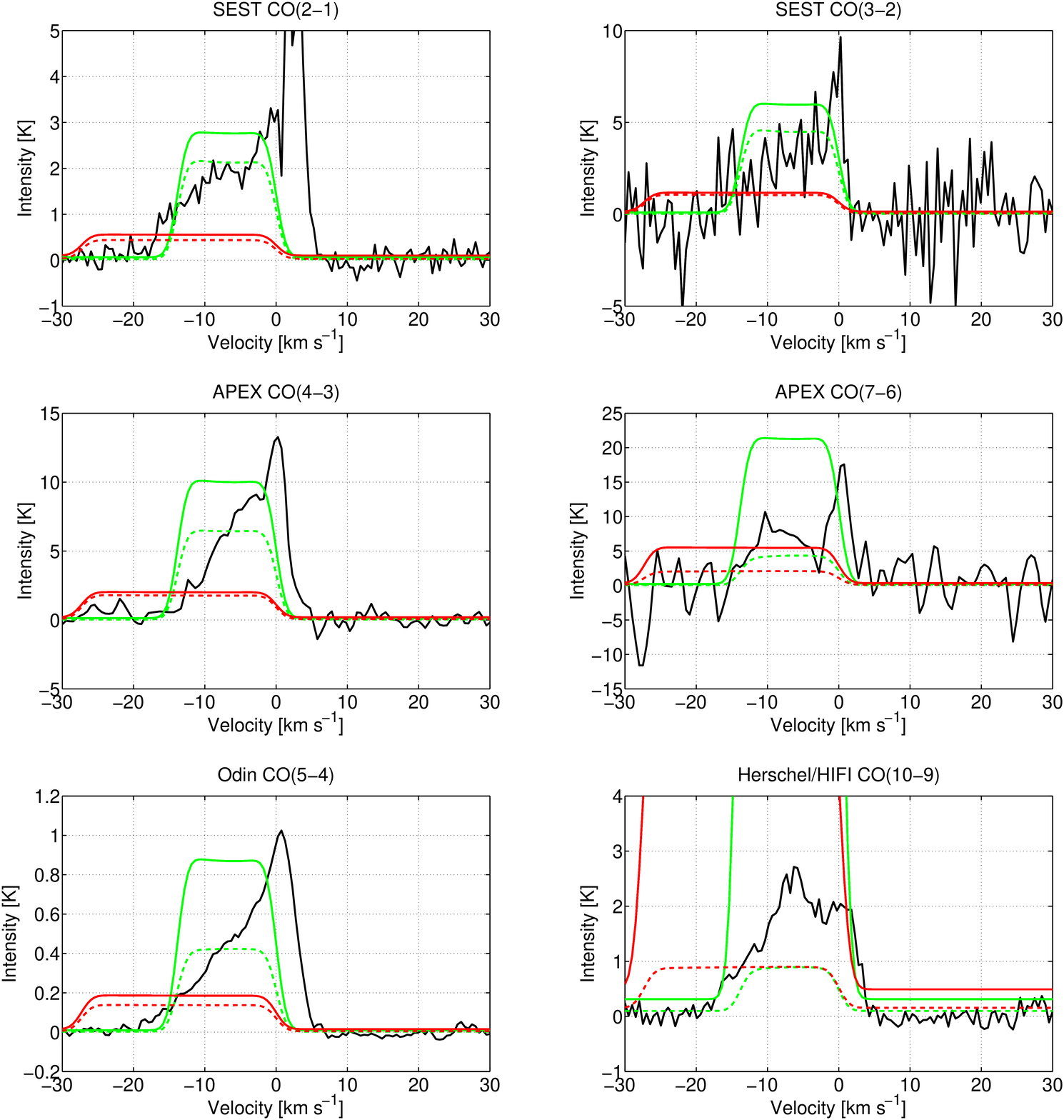}
     \caption{The red (1000~K) and green (400~K) lines represent the model described in \cite{Neufeld:2006fk}, using pre-shock densities of \powten{4}~\cmthree\ (dashed) and \powten{5}~\cmthree\ (solid). The observed spectra from \sest, \apex, \odin\ and \hifi\ are plotted with black solid lines. Note that the modelled spectra, not have been baselines subtracted.}
     \label{figure:neufeldCO}
\end{figure*}

The results are presented in Figs.~\ref{figure:neufeldCO} and
\ref{figure:neufeldwater}. As seen in the latter figure, the computed
line strength of the ground state line of \ohtvao\ for a temperature
of 400~K and pre-shock density of \powten{5}~\cmthree\ is not far from
what \hifi\ has observed. On the other hand, such parameters are not
in agreement with the observed \cotionio\ line, the intensity of which
is severely over-predicted. As expected,
the low-J lines \cotvaett\ and \cotretva\ are not very sensitive to the
changes in temperature from 400 to 1000~K. In all cases the
computed rectangular line shape is different from the observed profiles,
which have a pronounced triangular shape.

\subsection{Emission from a curved geometry}
\label{section:curved}
The spatial distribution of the gas does also influence the observed
line profiles \citep[see e.g.][]{Hartigan:1987fk}. For that reason we
also investigate a scenario where the emission originates from a
curved geometry. To implement this we use a simple model that mimics
an expanding shell with a diameter of 30\asec\ located at a distance
of 180 pc (see Fig~\ref{figure:modelsphere}). The interior of the spherical shell is empty, i.e. at the
temperature of cosmic background radiation field. 
This cold sphere
occupies 88\% of the radius of the sphere, based on a shell thickness
of 5\texpo{15}~cm. This value is in between the slab
thickness estimated by \citet{Liseau:1996fk} and the analytical
expression for the slab thickness at 180~K. The shell
thickness is also consistent with the cooling length estimated by
\cite{Kaufman:1996qy}. Using this method, we block out essentially all
the radiation originating from the opposite side of the sphere.

To compare with the observed line profiles, spectra are computed
viewing the curved surface from the front. This is supported by the absence of
detectable SiO emission which is often observed in molecular outflows
with relatively high velocities \citep[see e.g.][]{Nisini:2007fk}.
Shock modelling carried out by
\citet{Gusdorf:2008lr} suggests that sputtering is not very efficient
in the velocity regime below 25~\kmpers. Therefore, assuming that the
outflow in HH\,54 is observed from the front, and that it has a small
inclination angle with respect to the line of sight, the maximum
velocity of the molecular gas is likely lower than this.

Inspection of the CO and \htvao\ spectra show a maximum radial
velocity of \about20 \kmpers. This is also consistent with the
modelling carried out by \citet{Giannini:2006lr} who suggested a C+J
type shock with a maximum velocity of 18~\kmpers. The velocity in the
shell increases linearly from 0 to 20~\kmpers\ (see
Table~\ref{table:1}). The true velocity profile is most likely more
complicated. In a bow shock, the velocity component perpendicular to
the jet direction is expected to be smaller than the component
parallel to the jet direction. This makes the model somewhat simpler
than reality. The velocity field within the shocked region probably
also has a more complicated profile. Furthermore, we assume that the
emission in all lines stem from the same region of size 30\asec. The
bump-like feature discussed in Section~\ref{section:obsline},
indicating a deviation from a linear velocity profile, has not been
considered in the modelling presented here.
 The parameters used in the model
are summarised in Table \ref{table:1}.

We set up a grid where we vary the \htva\ density from \powten{4} to
\powten{8}~\cmthree\ and the kinetic temperature of the gas from 30 to
330~K. Thus, our model is steady state and in equilibrium. We choose this approach in order to keep the number of free parameters as small as possible. Also, it is worth noting, that we have not achieved a better fit to the bump-like feature (see Sec.~\ref{section:obsline}) when varying the density and temperature profiles over the shells. 
To find the best fit density and kinetic temperature, the reduced  
$\chi^2$ is minimised, where the difference between the observed and the modelled intensity is evaluated in each velocity bin
(see Fig.~\ref{figure:6}). 
The  \cotvaett, \cotretva, \cofemfyra\ and \cotionio\ lines are included in the $\chi^2$~-~minimisation whereas the \cofyratre\ and \cosjusex\ lines are exluded due to the contamination from the off position
and the high noise level of the \cosjusex\ observation. For all the CO
spectra, we only take the line wings into consideration, due to the
fact that emission from the surrounding cloud is clearly visible in
the \cotvaett\ spectra. 
\begin{figure*}[t]
\centering
   \includegraphics[width=14cm]{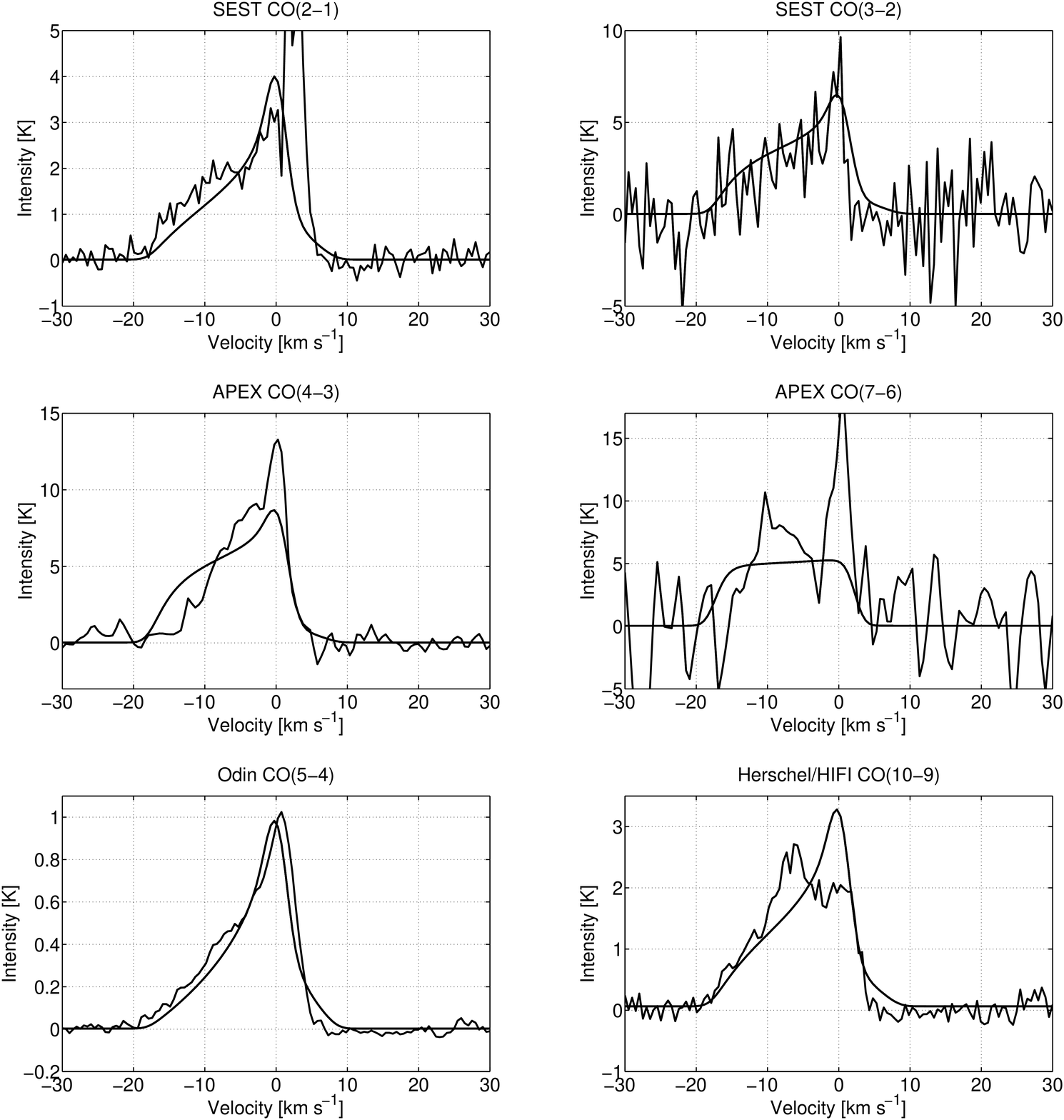}
     \caption{The six spectra obtained with \sest, \apex, \odin\ and \herschel\ compared to the best-fit model (see Sec.~\ref{section:abundance}). 
}
     \label{figure:1}
\end{figure*}

\subsubsection{Density, temperature and water abundance}
\label{section:abundance}
From the curved geometry model, the best fit gas density and kinetic
gas temperature are \nhtva~=~9\texpo{4}~\cmthree\ and \tkin~=~180~K
(formally 177~K) respectively. This implies a total \htva\ mass of \about 1\texpo{-2}~\msun. The value of the reduced $\chi^2$ is
2.3. The CO spectra
obtained with \odin, SEST and HIFI are plotted in Figure
\ref{figure:1} together with the modelled spectra. The model fits the
observations well, and we conclude that the geometry of the region can
be a crucial parameter determining the shape of the line profiles. The
\cotionio\ line observed with \herschel, however, shows a slightly
more complicated profile than predicted (see discussion in
Section~\ref{section:obsline}). Thus, the triangular form is distorted
by the presence of the \vlsr~$\simeq -7$~\kmpers\ feature. A minor
change in the kinetic temperature (i.e to 170~K) provides a better fit
to the underlying triangular shape of the \cotionio\ line without
affecting the lower-J CO lines by much. The computed maximum optical
depths are $\leq$7\texpo{-2} for all the modelled CO lines. Using the
gas density and kinetic temperature obtained from the CO modelling,
the observed \ohtvao\ ground state transition is best fit with an
\orthowater\ abundance with respect to \htva\ of
\xohtvao~=~1\texpo{-5} (see Figure \ref{figure:2}). 
  \begin{figure}[t]
\centering
   \includegraphics[width=8cm]{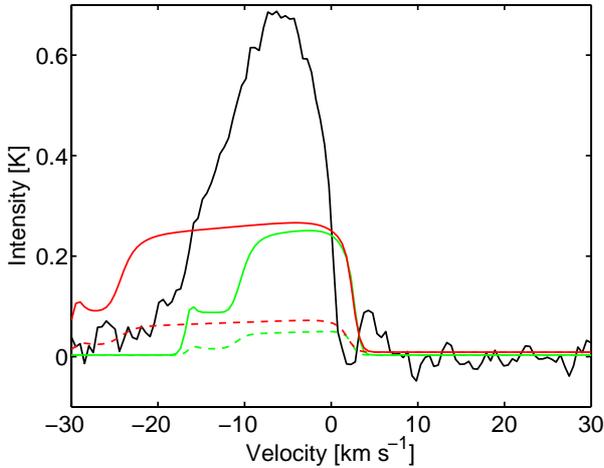}
     \caption{The red (1000~K) and green (400~K) lines represent the model described in \cite{Neufeld:2006fk}, using pre-shock densities of \powten{4}~\cmthree\ (dashed) and \powten{5}~\cmthree\ (solid). The observed spectrum for the \htvaoett\ transition with \hifi\ is plotted with a black solid line. The \ohtvao\ abundance is set to \xohtvao~=~1\texpo{-5}.}
     \label{figure:neufeldwater}
\end{figure}
\begin{figure}[t]
\centering
   \includegraphics[width=8cm]{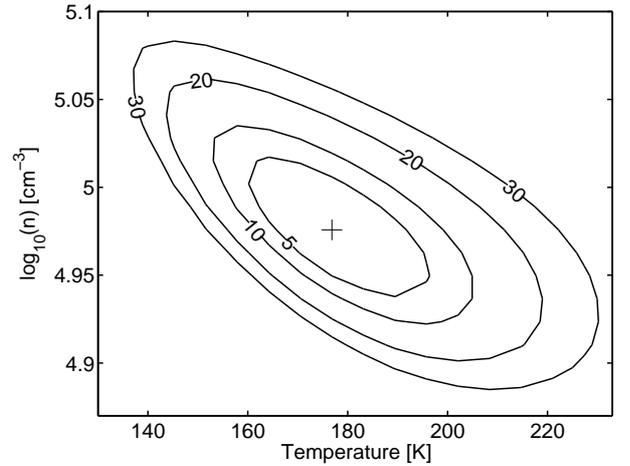}
     \caption{Contour plot showing the sum of the chi-squares, comparing the modelled spectra and the observations. The 5\%, 10\%, 20\% and 30\% deviations from the minimum value are indicated with solid lines. The best-fit density and temperature are indicated with a cross.}
     \label{figure:6}
\end{figure}
\begin{figure}[h]
\centering
   \includegraphics[width=8cm]{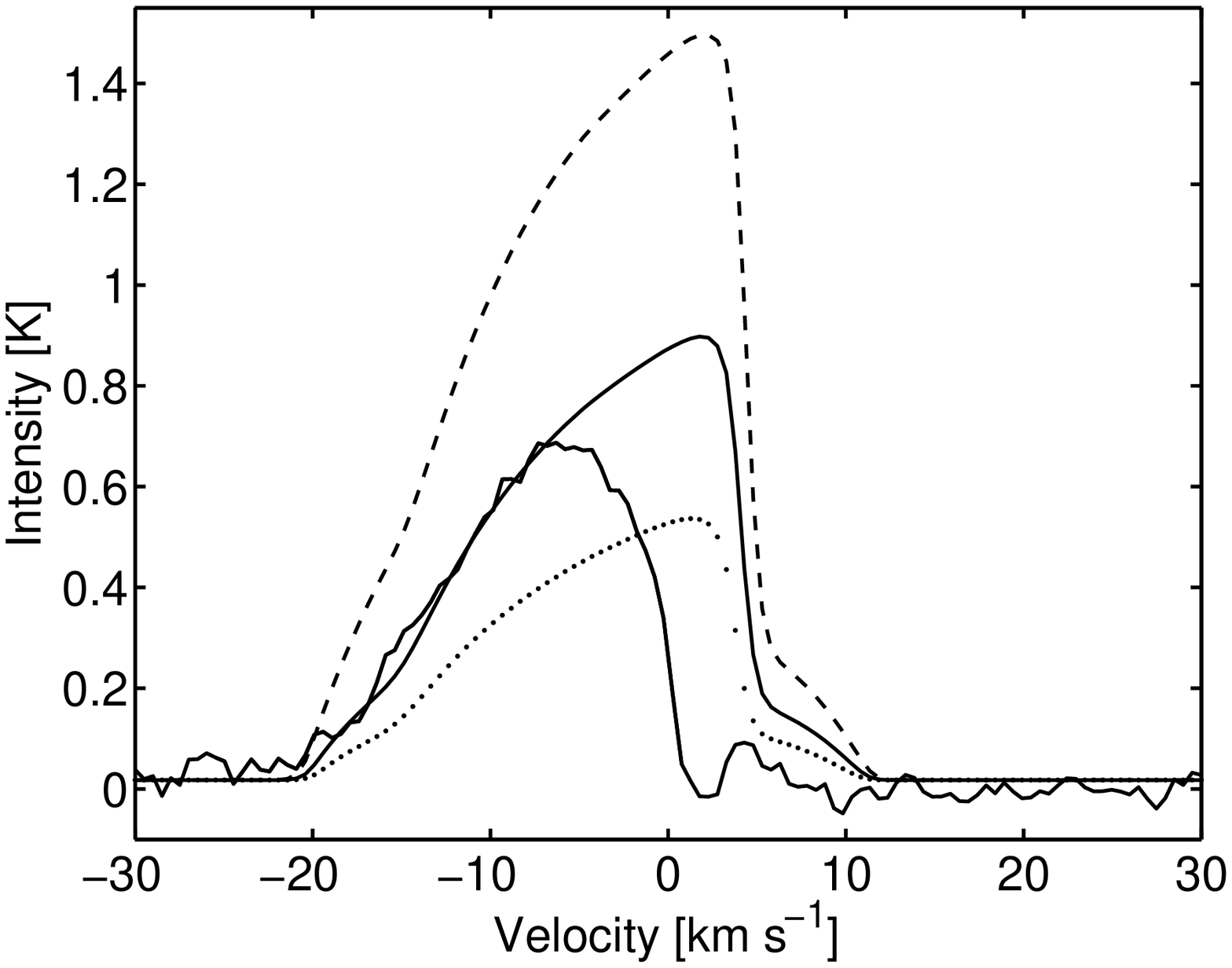}
     \caption{The \hifi\ \htvaoett\ spectrum compared with the model spectrum for \xohtvao~=~1\texpo{-5}. Spectra with the \ohtvao\ abundance increased by a factor of two (dashed line) and decreased by a factor of two (dotted line) are also included included. Note that the foreground gas is visible in absorption in the observed spectrum. 
     }
     \label{figure:2}
\end{figure}
The observed total flux in the map obtained with
PACS 
corresponds to an integrated line intensity of 20~K~\kmpers. This
is in agreement with the integrated intensity from the predicted line profile
to within a factor of 2.
\subsubsection{\htvao\ Line profile predictions}
\label{section:predictions}
\begin{figure}[t!]
\centering
   \includegraphics[width=8cm]{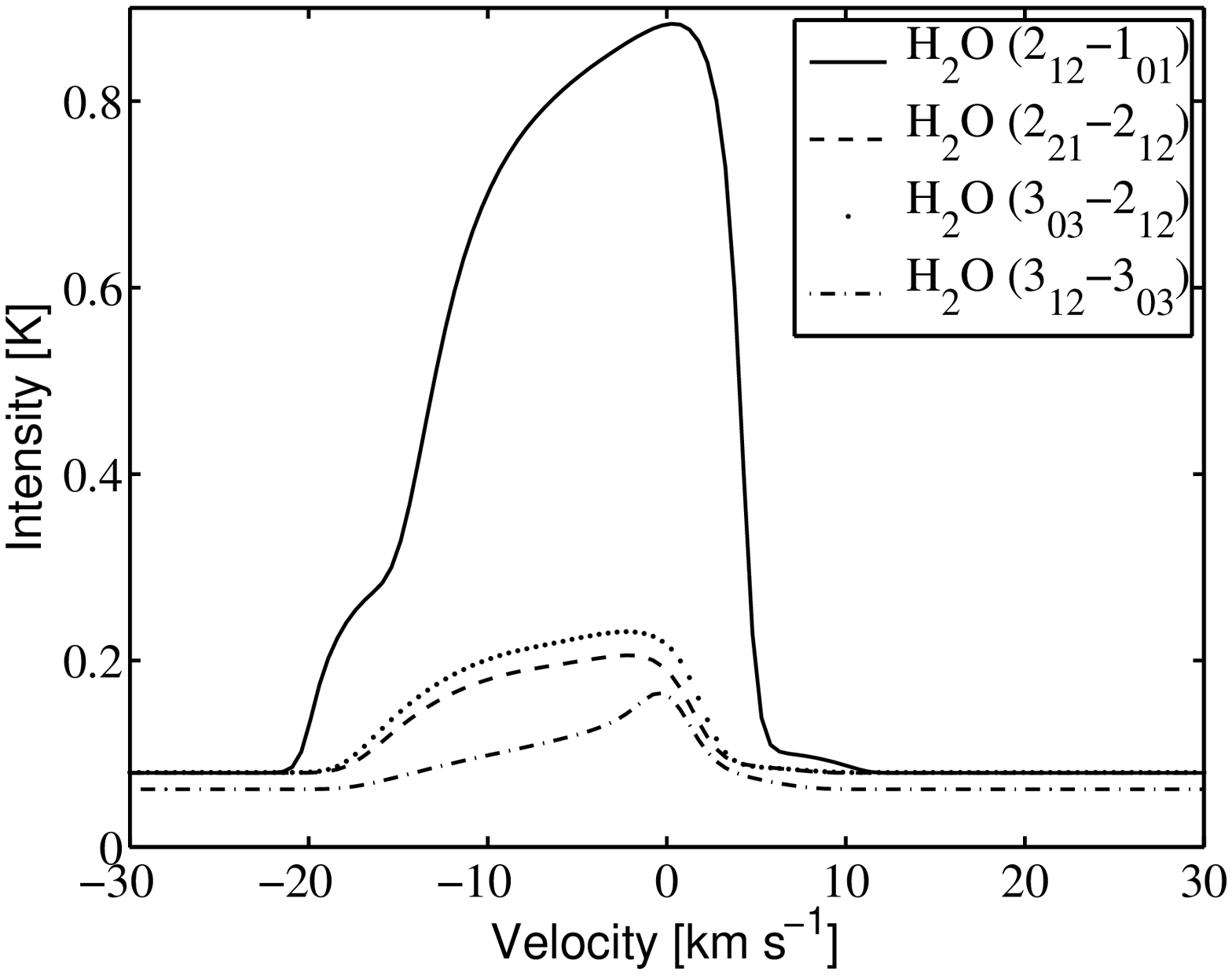}
     \caption{Line profile predictions for the \htvaotva, \htvaofyra, \htvaofem\ and \htvaosju\ spectra as observed with HIFI. 
}
     \label{figure:3}
\end{figure}

The line profiles for the six lowest rotational transitions of
\orthowater\ have been computed. The abundance is set to 1\texpo{-5}
and the lines, that are predicted to be strong enough to be readily detected with \hifi,
are displayed in Figure \ref{figure:3}.
Also in this figure, red-shifted emission originating from the
opposite side of the sphere is present for the model with a curved
geometry. 
Significant changes in the excitation temperature in the inner and
outer shells show up in these spectra as weak absorption features. For
this reason, the spectra have been computed using more than
100 shells to avoid any drastic changes in excitation temperature, due
to optical depth effects between each shell. Simultaneously with the
\cotionio\ observation, \htvaosex\ at 1153 GHz was also observed. The
observed noise level of \about0.1~K is however too high to detect the
line with a predicted strength of only \about10~mK. A summary of these
predictions is presented in Table~\ref{table:2}. In this table, the observed integrated intensity in the \htvaoett\ line is 30~\% lower than the predicted value due to the absorption from foreground gas. If this gas is at a low temperature, however, the higher transitions should not be affected by much.
\begin{table*}[t]
\caption{The predicted integrated intensities, maximum intensities, continuum levels and optical depths for the six lowest rotational transitions of \orthowater. The maximum and integrated intensities are for the baselines subtracted spectra. The foreground absorption, visible in the observed \htvaoett\ spectrum, is not considered in this table.
}
\label{table:2}
  \renewcommand{\footnoterule}{} 
  \begin{tabular}{lrrllllrr}
          \noalign{\smallskip}
         \hline
          \hline
 \noalign{\smallskip}
          Line & $\nu$ & FWHM & Receiver & $\int {T}_{\mathrm{mb}} d\upsilon$ & \expo{14}~$\int {F}_{\lambda} d\lambda$&  ${T}_{\mathrm{mb}, \mathrm{max}}$ & $T_{\rm cont}$ & $\tau_{\rm max}$  \\
          & (GHz) & (\asec) & & (K~\kmpers)& (erg cm$^{-2}$ s$^{-1}$)& (K) & (mK) & \\
          \noalign{\smallskip}
          \hline
          \noalign{\smallskip}
          \noalign{\smallskip}
          $\mathrm{1_{10}}$-$\mathrm{1_{01}}$ & 556.936 & 39 & HIFI & 12.9 &9.3& 0.88 & 18 & 40  \\
          \noalign{\smallskip}
          $\mathrm{2_{12}}$-$\mathrm{1_{01}}$ & 1669.905 & 13 & HIFI/PACS & 13.0 &28& 0.80 & 80 & 44  \\
          \noalign{\smallskip}
          $\mathrm{2_{21}}$-$\mathrm{1_{10}}$ & 2773.985 & 8 & PACS & 0.6 &2.4& 0.06 & 90& 3.9 \\
          \noalign{\smallskip}
          $\mathrm{2_{21}}$-$\mathrm{2_{12}}$ & 1661.017 & 13 & HIFI/PACS & 1.8 &3.8& 0.13 & 80 & 0.3 \\
          \noalign{\smallskip}
          $\mathrm{3_{03}}$-$\mathrm{2_{12}}$ & 1716.775 & 13  & HIFI/PACS & 2.3 &5.4& 0.15 & 80 & 0.5\\
          \noalign{\smallskip}
          $\mathrm{3_{12}}$-$\mathrm{2_{21}}$ & 1153.117 & 19 & HIFI & 0.1 &0.2& 0.01 & 65 & $<$0.1\\
          \noalign{\smallskip}
          $\mathrm{3_{12}}$-$\mathrm{3_{03}}$ & 1097.365 & 20 & HIFI & 1.0 &1.4& 0.10 & 62 & $<$0.1\\
          \noalign{\smallskip}
          $\mathrm{3_{21}}$-$\mathrm{2_{12}}$ & 3977.047 & 5 & PACS & 0.3 &1.3& 0.02 & 103 & 0.3\\
          \noalign{\smallskip}
          \noalign{\smallskip}
          \hline
 
  \end{tabular}
\end{table*}

\subsubsection{Cooling rate ratios}
\label{section:cooling}
For the curved model described in Sec.~\ref{section:curved}, we derive
the cooling ratio $\Lambda({\rm CO})/\Lambda$(\ohtvao) $ =10$ where
the 557~GHz line is the dominant contributor to the water
cooling. This value agrees well with the earlier determination of 7
based on ISO-LWS data and presented by \citet{Liseau:1996fk}. 

\subsubsection{The bump-like feature}
As already discussed in Sec.~\ref{section:obsline}, the bump-like feature seems unresolved to
\herschel, i.e the size of the source is uncertain.
A source size comparable to the
beam size of \herschel, at 1152~GHz (19\asec), would have to have a
high temperature ($>$1000~K) and low density
($\sim$\powten{3}~\cmthree) to fit the observations. This implies that
the flux in the high-J CO lines would be higher than what was observed
with ISO-LWS \citep{Giannini:2006lr}. However, a source size as small
as $\sim$1\asec, where the temperature and density would have to be
$\sim$400~K and $\sim$\powten{8}~\cmthree\ respectively, is a
plausible scenario. In that case however, the \orthowater\ abundance
has to be less than \powten{-8} to fit with the ISO-LWS
observations. In addition, the source is likely not smaller than
1\asec. Also in this case, the high-J CO lines would be stronger than
what is actually observed.

Using a source size of 10\asec\ (this size is estimated from \htva\
maps presented in \citet{Neufeld:2006fk}), the observed CO
$"$bullet$"$-emission is well fit with a temperature, \tkin~=~600~K,
and a density, \nhtva~=~2\texpo{4}~\cmthree. This yields a column
density, \Nhtva~=~5\texpo{20}~\cmtwo. For an \orthowater\ abundance of
\xohtvao~=~1\texpo{-5}, the \htvaofem\ emission is entirely accounted
for by the ISO-LWS observations. In this case the \htvaotva\ emission
would be merely about 25\% of what was observed with PACS and no
significant contribution would be observed in the \htvaoett\ line.
Therefore, we conclude that the \ohtvao\ abundance in the
$"$bullet$"$ is likely lower than
\powten{-5}. 
\subsubsection{Comparison with other results}
\label{section:lineprofiles}
The derived \htva\ density implies a column density of
5\texpo{20}~\cmtwo\ which is approximately one order of magnitude
higher than what is derived for the warm gas
\citep{Neufeld:2006fk}. The best fit temperature, \tkin~=~180~K, is
significantly lower than the temperature in the gas responsible for
the high-J CO emission [i.e CO\,(14$-$13) -- CO\,(20$-$19)] reported in
\citet{Giannini:2006lr}. Our model predicts these lines to be one to
two orders of magnitude weaker.  Clearly, different temperature regimes 
are present in HH\,54 and a one-temperature, one-density model cannot 
explain all the infrared observations.
\begin{figure}[]
\hspace{0cm}
   \includegraphics[width=9cm]{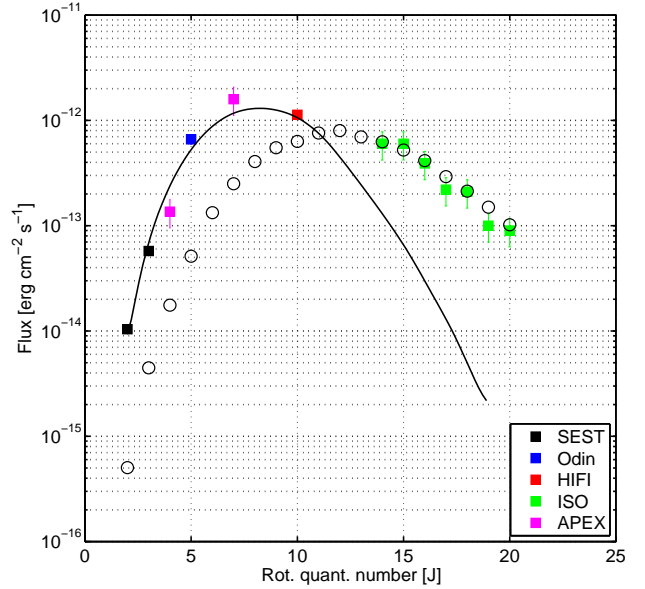}
   \caption{Observed CO line fluxes (squares) as a function of the
     rotational quantum number J. The errorbars are the estimated
     calibration uncertainties (10\% for \sest, \odin\ and \hifi\ and
     30\% for \apex\ and \iso). The best fit model, described in the
     present paper (see Sec.~\ref{section:abundance}), is indicated with a solid line. The fit to the high-J CO lines is indicated with black circles (see Sec.~\ref{section:lineprofiles}). 
      }
     \label{figure:fluxplot}
\end{figure}
The high-J CO emission observed with \iso\ can be explained using the
same geometry and source size but having a shell thickness
4\texpo{14}~cm, i.e one order of magnitude thinner. In that case a
slightly higher density (\nhtva~=~1.5\texpo{5}~\cmthree) and
temperature (T~=~500~K) fits the high-J CO lines well. On the other
hand, this secondary component also makes a significant contribution
to the \cotionio\ line and this emission may originate from different
components. Recently, \citet{Takami:2010uq} report morphological
differences between \spitzer\ observations in the 3.6, 4.5, 5.8 and
8.0~$\mu$m bands. The emission is observed to be less patchy in the
long wavelength bands (i.e. 5.8 and 8.0~$\mu$m) and they interpret
this as thermal \htva\ emission being more enhanced in regions of
lower density and temperature. Given the fact that hot gas obviously
is present in this region, the secondary component may in reality be in
smaller regions of high temperature similar to those observed in
\htva. In Figure~\ref{figure:fluxplot}, the CO line flux is plotted as
a function of the rotational quantum number, J.

The predicted integrated intensities can be compared with the
modelling presented in \cite{Giannini:2006lr}. These authors present a
multi-species analysis where they conclude that the observed \htva, CO
and \htvao\ lines can only be explained by a J-shock with magnetic
precursor. The C+J shock model, presented in their paper, explains the
observed \htvaotva\ and \htvaofem\ line fluxes of 7\texpo{-13}~\ecs\
and 2\texpo{-13}~\ecs\ well, and they predict the line flux for the
\htvaoett\ line to be 2.0~\texpo{-13}~\ecs. Converted to a K~\kmpers\
scale these values correspond to 6.8, 1.4 and 53~K~\kmpers\
respectively. Taking the beam size into account, the predicted
integrated intensity for the \htvaoett\ line in this paper is at least
a factor of four lower. This could be due
to the fact that this line is very sensitive to the type of shock
present. This is also discussed in the paper by
\cite{Giannini:2006lr}, where they note that a J type shock would
change the flux by more than a factor of two downwards.

Recently, \citet{Flower:2010fkq} presented theoretical predictions of CO
and \htvao\ line intensities, based on detailed C- and J-type shock
model calculations. Expectedly, the assumed magnetic fields are
different for their models of C- and J-shocks, i.e. for $b=1$ and
$b=0.1$, respectively, where $b$ is defined through $B=b\sqrt{n_{\rm
    H}}\,\,\mu$G. For their high density C-shocks (see below), this
means that pre-shock fields of order 450\,$\mu$G should be
present. Based on OH Zeeman observations, \citet{Troland:2008kx} detected
9 dark clouds in a sample of 34. Corresponding line-of-sight magnetic
field strengths were in the range $B_{\rm los} =10-25\,\mu$G, with
typical values around 15\,$\mu$G. As discussed by \citet{Troland:2008kx},
for randomly oriented fields, $B_{\rm los} = 0.5 \times
\!\!\mid\!\!{\bf B}\!\!\mid$. Hence, magnetic fields in dark clouds do
not likely exceed levels of $20-50\,\mu$G. This refers to the observed
scales of 3\amin. However, these authors also showed that in a given
cloud, $B_{\rm los}$ did not change appreciably from one position
(active molecular outflow) to another (quiescent surrounding
cloud). It seems, therefore, that fields as strong as $450\,\mu$G may
not be that common. On the other hand, order of magnitude lower field
strengths ($45\,\mu$G for $b=0.1$) are more consistent with the
observational evidence and may promote the occurrence of J-shocks.

For the radiative transfer \citet{Flower:2010fkq} used an LVG
approximation in slab geometry. In their Fig.\,8, line profiles for
CO\,(5-4) and \htvaoett\ are shown for C-shocks with two densities and
four shock velocities. These lines are close in frequency and
observations of HH\,54 with Odin are made with essentially the same
telescope beam, rendering resolution issues to be of only minor
importance, i.e. any scaling of the intensities due to the source size
should affect both observed lines in the same way. It should be
feasible, therefore, to directly compare the line profiles of our Odin
observations with those of the models by \citet{Flower:2010fkq}.
Based on the observed maximum radial velocities,
we consider only models with $\upsilon_{\rm s} \ge 20$\,\kmpers.  Their
models with shock velocities of 20\,\kmpers\ could correspond to a
head-on view, whereas their models for \vs\,=\,30 and 40\,\kmpers\ could
correspond to inclinations of the flow with respect to the line of
sight of 48\adeg\ and 60\adeg, respectively\footnote{For the emission
  knots of HH\,54 \citet{Caratti-o-Garatti:2006lr} determined an average
  inclination of 27\adeg, which would imply that $\upsilon_{\rm s}
  \sim \upsilon_{\rm obs}$.}.

For the C-shock models, several of the line profiles display shapes
that are qualitatively similar to those observed with \odin\ and
\hifi. These line shapes are a consequence of the computed flow
variables, not a geometry effect. However, in particular for the
range of $\Delta \upsilon \sim 0$ to about 10\,\kms, the \odin/\hifi\
lines exhibit CO-to-\htvao\ intensity ratios very much in excess of
unity, i.e. $T_{\rm CO}(\Delta \upsilon)/T_{\rm H_2O} (\Delta
\upsilon) \gg 1$. $\Delta \upsilon$ is the velocity relative to the
rest frame of the flow, i.e. \vlsr.

\begin{figure}[]
\centering
\includegraphics[width=8.5cm]{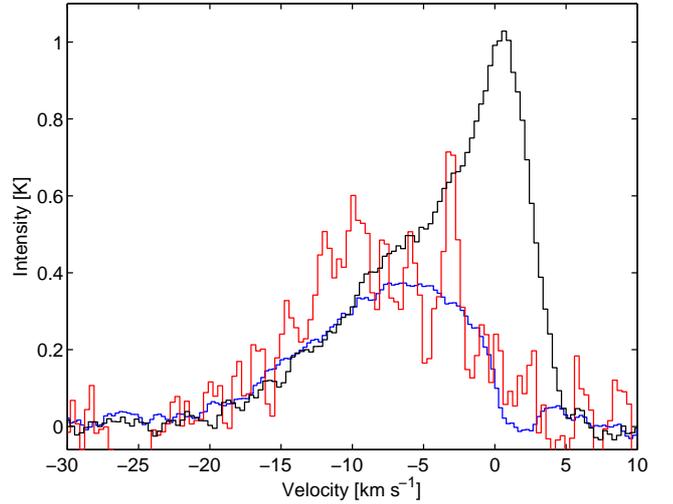}
   \caption{Comparison between the \cofemfyra\ line (black) and the \htvaoett\ line obtained with \odin\ (red). The \htvao\ spectrum has been
     multiplied with 6 for clarity. Also the \hifi\ \htvaoett\ spectrum has been plotted in this figure for comparison (blue). The intensity of the emission has in this case been multiplied by 6 and corrected for beam filling.       }
     \label{figure:flowercomp2}
\end{figure}

\begin{figure}[]
\centering
\vspace{-0.3cm}
   \includegraphics[width=8.8cm]{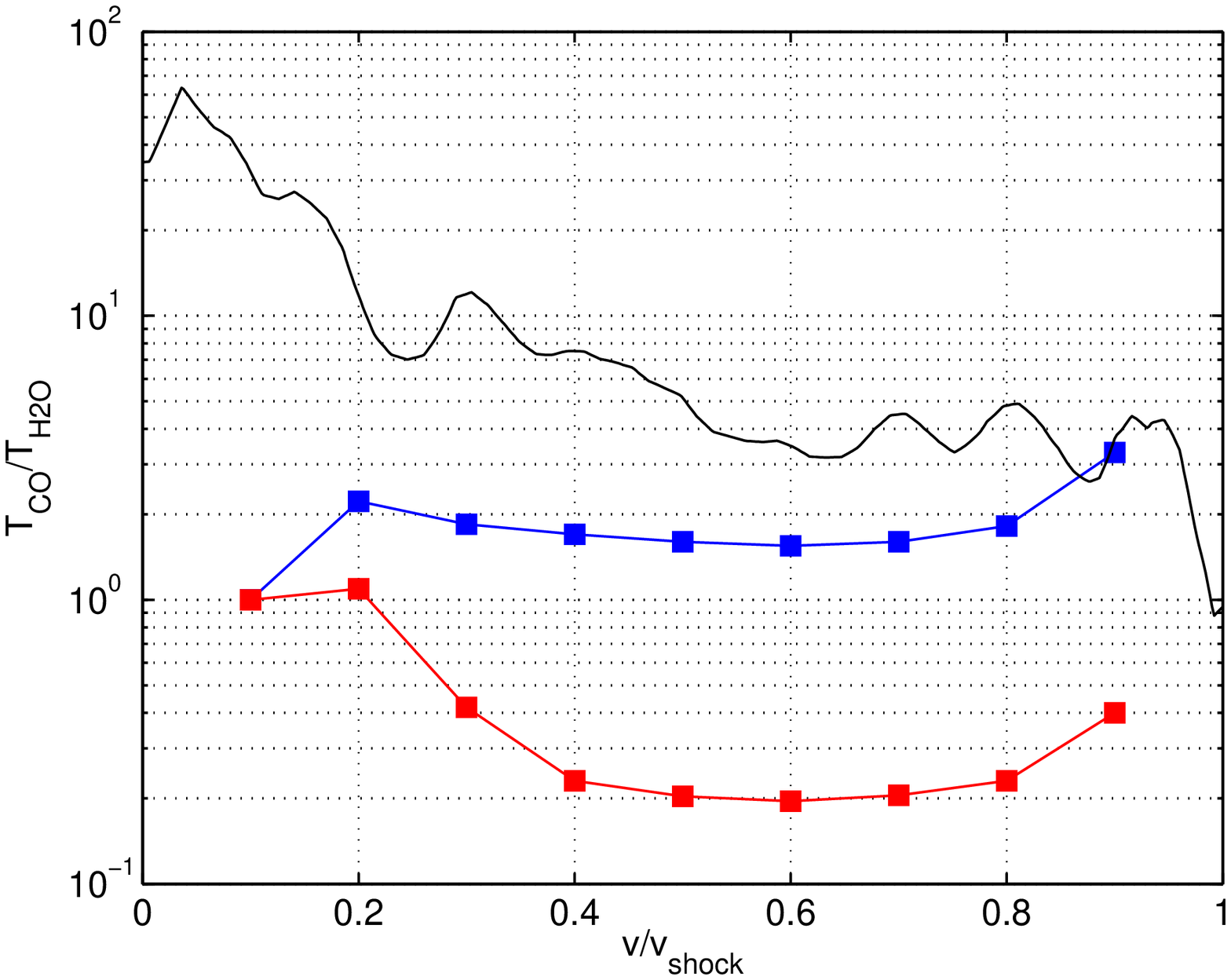}
   \caption{Line ratios between \cofemfyra\ and \htvaoett. The ratio as
     measured from \odin\ is indicated with a black line. 
     The predictions presented by \citet{Flower:2010fkq} for
     $\upsilon_{\mathrm{shock}}$~=20~\kmpers\ are plotted in blue for
     $n$(H)~=~2\texpo{4}~\cmthree\ and in red for
     $n$(H)~=~2\texpo{5}~\cmthree.}
     \label{figure:flowercomp}
\end{figure}

These observed ratios are very much larger than the theoretical values
(Fig.\,\ref{figure:flowercomp2} and \ref{figure:flowercomp}). 
Albeit the ratios are larger than
unity for the low-density ($n_{\rm H}=2\times 10^4$\,\cmthree)
C-shocks, these fall still far below the observed ones. On the other
hand, the high density ($n_{\rm H}=2\times 10^5$\,\cmthree) cases
could directly be dismissed, as these tend to show {\it inverted}
ratios, i.e. $T_{\rm CO}/T_{\rm H_2O}\,\ll 1$, contrary to what is
observed (Fig.\,\ref{figure:flowercomp}). 

For the J-shock models, \citet{Flower:2010fkq} list the predicted
integrated intensities. Also in this case, the predicted CO-to-\htvao\
ratios (for a shock velocity of 20\,\kmpers) are much lower than
the observed ratio, viz 0.06 and 0.005 for the pre-shock densities
$n_{\rm H}=2\times 10^4$\,\cmthree and $n_{\rm H}=2\times
10^5$\,\cmthree\ respectively.
\subsection{Implications for future work}
The modelling of planar C-shocks show that low densities are
required for cooling rate ratios, $\Lambda({\rm
  CO})/\Lambda$(\ohtvao)~$> 1$, in contrast to our own findings, where
densities at least as high as \powten{5}\,\cmthree\ seemed implied by
the observations (Sect.~\ref{section:abundance}). The cause for this
mismatch is not clear to us, but one of the reasons could be the
difference between modelled planar and curved geometry. Such modelling
should therefore be attempted.  The relatively high cooling rate
ratio, $\Lambda({\rm CO})/\Lambda$(\ohtvao)$~=~10$, is also not easily
reconcilable with the presence of a C-type shock where a ratio,
$\Lambda({\rm CO})/\Lambda$(\ohtvao)~$\ll 1$ is favoured. 
On the other hand, J-shock models predict ratios even smaller than for C-shock models. However, the relatively low \orthowater\ abundance determined from the modeling indicate that J-type shocks may contribute.

As already discussed in Sect.~\ref{section:lineprofiles}, different
temperature regimes are present in the HH\,54 region. It would be
adequate therefore, to repeat the ISO-LWS observations, using the
higher sensitivity and spatial resolution provided by PACS. The mid-J
CO lines fall in the wavelength range covered by the Spectral and
Photometric Imaging Receiver (SPIRE) and these should be observed.

\section{Conclusions}

Based on spectral mapping with Herschel of the region containing the Herbig-Haro objects HH\,52 to HH\,54 we conclude the following:

\begin{itemize}
\item[$\bullet$] The CO\,($10-9$) 1152\,GHz line was clearly detected
  only toward the position of HH\,54 
  with a FWHM\,\lapprox\,30\asec, comparable
  to the extent of the HH-emission knots in the visible and infrared.
  \item[$\bullet$] The \htvaotva\,1669\,GHz line was clearly detected
  toward HH\,54 and with a similar extent as the CO\,($10-9$) line. An
  offset of 9\asec\ ($2.4 \times 10^{16}$\,cm) between the two species
  is observed, but the reality of this can at present not be firmly
  assessed. The \htvaoett\,557~GHz line was also clearly detected
  toward HH\,54.
\item[$\bullet$] The CO\,($10-9$) spectra show only blueshifted
  emission, with maximum relative velocities in excess of
  $-20$\,\kmpers. The line profiles exhibit typically a triangular
  shape, which in certain positions is however contaminated by a
  bump-like feature at \vlsr\,$= -7$\,\kmpers. This feature is
  constant in velocity and width. It is limited in spatial extent and may be
  identified by what is commonly called a $"$bullet$"$. Comparison of
  the observed spectra with analytical bow shock line profiles limits
  the viewing angle to greater than zero but \lapprox\,30\adeg.
\item[$\bullet$] The bump is clearly seen in position-velocity cuts,
  revealing two peaks, in addition to a smooth velocity gradient of
  about \powten{3}\,\kmpers\,pc$^{-1}$. The
  low-velocity peak appears close to the ambient cloud velocity,
  whereas the second peak corresponds to the bump. For this feature we
  determine from Gaussian fit measurement of the CO\,($10-9$) data a
  relative peak-$T_{\rm mb}=1.5$\,K, \vlsr\,$=-7$\,\kmpers\ and
  FWHM\,=\,5\,\kmpers. 
  \item[$\bullet$] These line features are also observed in the
  CO\,($5-4$) spectrum observed with Odin and in spectra obtained from
  the ground, i.e. in maps of the ($2-1$), ($3-2$), ($4-3$) and
  ($7-6$) CO transitions. In addition to the triangular line shape,
  these data do therefore also confirm the reality of the spectral
  bump.
\item[$\bullet$] We initially use physical parameters for shock models
  of HH\,54 found in the literature to compute the CO spectra. These
  models presented two-component fits, given by analytical expressions
  for temperature and density in the respective slabs. These models
  were only moderately successful in reproducing the observational
  results, in particular what regards the line shapes. Considerable
  improvement was found, however, in computed spectra using a curved
  geometry instead. 
\item[$\bullet$] Using the best fit model parameters (in a
  $\chi^2$-sense) for the CO data we computed spectra for \htvao. This
  model fits the \htvao\,557\,GHz line observed
  with \hifi\ for an ortho-water abundance with respect to molecular
  hydrogen, $X({\rm H_2O})= 1 \times 10^{-5}$.
  \item[$\bullet$] A cooling rate ratio, $\Lambda({\rm
    CO})/\Lambda$(\ohtvao)~$\gg 1$, is not easily reconcilable with recent shock modelling. 
   On the other hand, a relatively low water abundance ($\sim$\expo{-5}), supports a scenario where J-shocks
  contribute significantly to the observed emission. This is also
  consistent with magnetic field strengths measurements toward dark
  clouds, where $B$-values lower than what is needed for C-type shocks,
  typically are observed.
\item[$\bullet$] Comparison of our \odin\ data with line profiles from
  recent detailed C-shock model computations in the literature
  suggests that some refinements of these models may be required.
\end{itemize}

\begin{acknowledgements}
The authors appreciate the support from A. Caratti o Garrati for
providing the \halpha\ image shown in
Figure~\ref{figure:0}. Aa. Sandqvist is thanked for scheduling the
\odin\ observations of HH\,54. We also thank the WISH internal referees Tim van Kempen and Claudio Codella for their efforts.\\ \\
HIFI has been designed and built by a consortium of
institutes and university departments from across Europe, Canada and the
United States under the leadership of SRON Netherlands Institute for Space
Research, Groningen, The Netherlands and with major contributions from
Germany, France and the US. Consortium members are: Canada: CSA,
U.Waterloo; France: CESR, LAB, LERMA, IRAM; Germany: KOSMA,
MPIfR, MPS; Ireland, NUI Maynooth; Italy: ASI, IFSI-INAF, Osservatorio
Astrofisico di Arcetri- INAF; Netherlands: SRON, TUD; Poland: CAMK, CBK;
Spain: Observatorio Astronomico Nacional (IGN), Centro de Astrobiologia
(CSIC-INTA). Sweden: Chalmers University of Technology - MC2, RSS \&
GARD; Onsala Space Observatory; Swedish National Space Board, Stockholm
University - Stockholm Observatory; Switzerland: ETH Zurich, FHNW; USA:
Caltech, JPL, NHSC.

\end{acknowledgements}
\newpage
\bibliographystyle{aa} 
\bibliography{papers}

\newpage
\newpage
\begin{appendix}
\section{}
\subsection{Extinction by dust and dust parameters}
\label{section:dust}
At the position of HH\,54, the molecular cloud is relatively tenuous
as the dust extinction through the cloud amounts merely to
\av\,\about\,2-3\,mag \citep{Gregorio-Hetem:1989fk,Cambresy:1999uq}. With the
relations of Eqs\,(2) and (8) of \citet{Hayakawa:2001qf}, this results in
a total \molh-column density of \Nhtva=$2\times 10^{21}$\,\cmtwo.

From [Fe\,II] line observations, \citet{Gredel:1994lr} determined the
visual extinction toward the HH object as \av\,=\,1-3\,mag, which
could suggest that the blueshifted HH\,54 is located at the rear side
of the cloud, a circumstance that potentially explains the apparent
absence of redshifted emission\footnote{From \molh-line emission at
  various positions of the HH object, \citet{Giannini:2006lr} derived the
  essentially constant value of \av\,=\,1\,mag.}. However, an $R_{\rm
  V}$-value of \gapprox\,5, which is significantly higher than that
used by \citet{Gredel:1994lr} and by others, needs to be invoked to
describe the Cha\,II dust \citep[see the discussion by][and references
therein]{Alcala:2008uq}. This reflects a predominantly 'big-grain'
contribution to the extinction (almost 2 magnitudes larger \av\ per
unit \ebv) and we consequently chose a shallower dust opacity
dependence on the wavelength in the FIR and submm, identified by the
parameter $\beta$ in Table\,3, i.e. $\beta=1$ as opposed to $\beta=2$
for the general ISM 
For the normalisation wavelength used by \citet{Hildebrand:1983fj},
i.e.  $\lambda_0=250$\,\um, we adopt the grain opacity
$\kappa_0=25$\,cm$^2$\,g$^{-1}$ \citep{Ossenkopf:1994fk}.
 The corresponding mass absorption
coefficient $\kappa$ at long wavelengths is then readily found from
$\kappa_{\lambda}=\kappa_0 (\lambda_0/\lambda)^{\beta}$.

Combining our own results with those found in the literature 
yields an SED of a putative core at 10\,K which provides a strict upper limit to the luminosity. Any associated central point source would have a limiting luminosity of $L_{\rm bol} <
0.4$\,\lsun.  Assuming optically thin emission at long wavelengths and
using standard techniques, the mass of the 10\,K core/envelope is
estimated at a mere 2\,\mearth, excluding the possibility of a very
young age of the hypothetical central source.

\subsection{The exciting source}
\label{section:source}
Several IRAS point sources are found in the surroundings of HH\,54 and
it is plausible to assume that one of these would be the exciting
source of the HH object. Most often, the exciting source is a young
(\lapprox\,Myr) stellar object, situated at the geometrical centre of
the bipolar outflow and this seems also to be the case for the CO
outflow associable with IRAS\,$12515-7641$\footnote{\citet{Alcala:2008uq}
  identify this source as a likely background K-giant.} near HH\,52
and HH\,53 \citep{Knee:1992lr}. For HH\,54, the situation is different
however: \citet{Knee:1992lr} identified IRAS\,$12522-7640$, which is
positionally coincident with HH\,54, as the driving source of a
collimated blueshifted outflow. It seems however very unlikely that
this IRAS source is (proto-)stellar in nature.  It was detected
essentially only, and barely, in the 60\,\um-band, and it was shown
that this IRAS flux is entirely attributable to \oishort\ line
emission \citep{Liseau:1996fk}. More recent observations by {\it Spitzer}
are in support of this conclusion, as these were unable to reveal any
point source at this position \citep{Alcala:2008uq}.

On the basis of proper motion measurements and morphology
considerations, \citet{Caratti-o-Garatti:2009fk} proposed the Class\,I
object IRAS\,$12500-7658$ to be the exciting source of HH\,54. The
IRAS source is located 20\amin\ south of HH\,54, i.e. at the projected
distance of 1\,pc. It was not detected at 1200\,\um\ by \citet[][see
below]{Young:2005kx}. What regards HH\,52 and HH\,53, no satisfactory
candidate was found by these authors. The possibility of different
exciting sources was also considered by \citet{Nisini:1996fk} and
\citet{Alcala:2008uq}, associating HH\,52 and HH\,53 with
IRAS\,$12496-7650$ (DK\,Cha), whereas \citet{Ybarra:2009zr} identify
this object also with the exciting source of HH\,54.

In APEX-LABOCA maps at 870\,\um, no emission was detected toward
HH\,54 nor from the Cha\,II cloud itself \citep[$S_{870\mu {\rm m}} <
20$\,mJy beam$^{-1}$:][]{van-Kempen:2008kx}. However, {\it jet-like}
extensions are seen to emanate from both DK\,Cha and
IRAS\,$12553-7651$ (ISO-Cha\,II\,28), which are situated some 10\amin\
to 20\amin\ south/southeast of HH\,54. Narrow, extended emission from
these IRAS sources was also reported by \citet{Young:2005kx}, who used
SEST-SIMBA to map the region at 1200\,\um. The alignment with the HH
objects is poor, however. Based on the observed spectral indices, this
emission is dominated by optically thin radiation from dust. The dust
features appear as tori or rings, seen edge-on, but given their sizes
($3\times 10^4$\,AU), they do not represent what commonly is called a
(protostellar) ``disc''. Perhaps, these dust features are left-overs
from an earlier history, during which the IRAS sources were formed. A
``conventional disc'' of size \about\,\powten{2}\,AU could very well
hide inside and any outflow would be orthogonal to these disc
features.

Whereas the dust extension from DK\,Cha is not, whether direct or
perpendicular, pointing anywhere near HH\,52-54, the vector orthogonal
to the linear feature of ISO-Cha\,II\,28, at position angle 51\adeg,
is within 5\adeg\ from the current direction toward HH\,54. Taking the
uncertainty of this estimate into account, this coincidence appears
very compelling. If this source drives/has driven a jet, that would
very well be aligned with HH\,54. The projected distance is 16\amin\
(0.8\,pc), i.e. jet travel times would be of the order of
$10^4\,(\upsilon/100\,{\rm km\,s}^{-1})^{-1}$\,yr. With a total
luminosity of \about\,10\,\lsun\ and a mass accretion rate of
\mdot$_{\rm acc} = 6 \times 10^{-7}$\,\msunyr\ \citep{Alcala:2008uq},
ISO-Cha\,II\,28 would make an excellent candidate\footnote{From the
  data for CO outflows compiled by \citet{Wu:2004fk}, and
  complementing information in the literature, it is found that
  $L_{\rm bol} \propto$\,\mdot$_{\rm loss}^{\,\,\,x}$, where $x=1.0$
  for \mdot$_{\rm loss} < 10^{-5}$\,\msunyr\ (for $\upsilon_{\rm
    wind}=$\,100\,\kms) and which steepens to $x=1.2$ beyond
  that. This holds over eight orders of magnitude, albeit with
  considerable spread due to the inhomogeneity of the data. According
  to this relation, $L_{\rm bol}=10$\,\lsun\ if \mdot$_{\rm loss}=
  4\times 10^{-7}$\,\msunyr.} for the exciting source of HH\,54, the
(distance corrected) stellar mass loss rate for which has been
determined as \mdot$_{\rm loss} \le 4 \times 10^{-7}$\,\msunyr\
\citep{Knee:1992lr}, assuming a flow velocity of $\ge100$\,\kmpers. A
ratio of loss-to-accretion rate of $\le 2/3$ would thus be indicated,
a value which is in reasonable agreement with theoretical expectation
for centrifugally driven MHD winds
\citep{Konigl:2000ys,Shu:2000vn,Contopoulos:2001bh}. To summarise, as
also concluded by others before, the issue of the exciting source(s)
of the HH objects 52, 53, 54 remains essentially
unsettled. 
\section{HIFI data reduction}
\label{section:reduction}
The \cotionio\ data published in this paper are public to the scientific
community and can be downloaded from the Herschel Science
Archive\footnote{http://herschel.esac.esa.int/Science\_Archive.shtml}
(HSA). The two \cotionio\ observations were carried out in
dual-beam-switch raster mode (observation id: 1342180798) and
on-the-fly with position switch mode (observation id: 1342190901). The
\htvaoett\ observation was carried out in point mode with position
switch. The data reduction and production of the \hifi\ maps followed
the following steps:
\begin{enumerate}
\item The raw data were imported to HIPE from the HSA.
\item All data were re-calibrated using v4.2 of the \hifi\ pipeline
  for \cotionio\ and v5.0 for \htvaoett. There is a slight offset
  between the pointings for the horisontal and vertical
  polarisations. For that reason, separate readout positions were
  calculated for the two polarisations.
\item The level 2 data were exported to Class using the HiClass tool in
  HIPE.
\item A baseline was fitted to and subtracted from each
  individual spectrum. This step and the subsequent data reduction was
  carried out in xs\footnote{Data reduction software developed by
    P. Bergman at the Onsala Space Observatory, Sweden;
    http://www.chalmers.se/rss/oso-en/observations/}. 
\item From Figure~\ref{figure:0}, it is clear that the readout
  positions for the two \cotionio\ maps are not on a regular grid in RA
  and Dec. Instead of averaging nearby spectra together, a Gaussian
  weighting procedure was used. The FWHM of the two dimensional
  Gaussian used in the weighting was taken to be 9.3\asec. This size
  is larger than the average spacing between individual spectra but on
  the other hand small enough not to change the spatial resolution
  significantly.
\item In order to verify the quality of the \cotionio\ observations,
  spectra taken toward the same positions in the two different
  observing modes were compared, showing no significant
  differences. (see Figure~\ref{figure:obscomp}). Also the horisontal
  and vertical polarisation data were reduced individually for both
  \cotionio\ and \htvaoett\ showing excellent agreement.
\begin{figure}[]
  \resizebox{\hsize}{!}{
  \rotatebox{270}{\includegraphics[width=9cm]{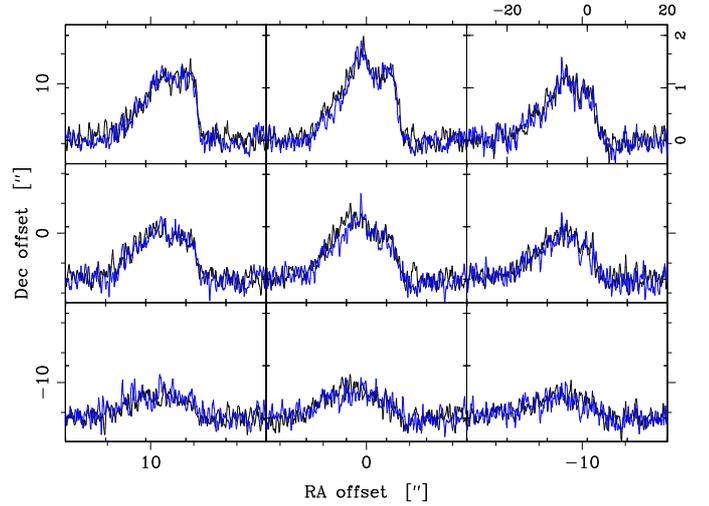}}
  }
  \caption{\cotionio\ spectra toward HH\,54. The black spectra are from the \hifi\ observation 1342180798, while the blue spectra are from the observation 1342190901. The horisontal and vertical polarisations have been averaged together (see text). }
  \label{figure:obscomp}
\end{figure} 
\end{enumerate}

\section{Radiative transfer analysis}
\label{section:radtrans}
An Accelerated Lambda Iteration (ALI) method
\cite[e.g.][]{Rybicki:1991vl} is used to compute the line profiles for
the observed transitions, varying only the density, kinetic
temperature and the geometry of the shocked region. ALI is a method, where
the coupled problem of radiative transfer and statistical equilibrium
is solved exactly. 
\subsection{Accelerated Lambda Iteration}
\label{section:ALI}
The ALI
code has in recent years been used in several publications \citep[see
e.g][]{Justtanont:2005lr,Wirstrom:2010uq} and was benchmarked with
other codes in a paper by \citet{Maercker:2008gd}. In that paper, the
ALI technique is described in more detail. In the present work,
typically 30 shells and 16 angles are used.
The statistical equilibrium equations
and the equation of radiative transfer are solved iteratively until
the relative level populations between each iteration changes by less than
\powten{-8}. In the modelling we set the energy limit to 2000~K. 
27 rotational energy levels, 26 radiative transitions and 351
collisional transitions are included in the calculations for CO. For
\ohtvao, the corresponding numbers are 45, 164 and 990. \subsubsection{Collision  rates}
In recent years, substantial efforts have been made to determine
cross-sections for \htvao\ excitation due to collisions with
\htva. Quantal calculations governing collisions between \htvao\ and
\htva\ are to this date not complete and for \ohtvao, only collision
rates for interaction with \phtva\ are available \citep{Dubernet:2009lr}. 
For this reason we make the assumption that
all \phtva\ are in the lowest energy state when solving the
statistical equilibrium equations. It is also assumed that the \phtva\
stays in the ground state after the collision. The ortho-to-para ratio
has been estimated for warm \htva\ by \citet{Neufeld:2006fk}. These
authors derive ratios significantly lower than 3, viz \about0.4 - 2 in
the HH\,54 region. The computed spectra from the ALI modelling have also been
compared with those, using the collision rates presented by
\citet{Faure:2007lr}, and with similar results. In the case of CO, we use the collision rates
presented by \citet{Yang:2010vn}.

\end{appendix}

\end{document}